\documentclass[prb,twocolumn,showpacs,preprintnumbers,amsmath,amssymb,floatfix]{revtex4-1}
\usepackage{graphicx}
\usepackage{bm}

\newcommand \be{\begin{eqnarray}}
\newcommand \ee{\end{eqnarray}} 
\newcommand \ba{\begin{align}}
\newcommand \eea{\end{align}}

\newcommand {\p}[1]{\partial_{#1}}

\newcommand \V{\vec}

\begin{document}
           \csname @twocolumnfalse\endcsname
\title{Dynamical charge and pseudospin currents in
graphene and possible Cooper pair formation}
\author{K. Morawetz$^{1,2,3}$ 
}
\affiliation{$^1$M\"unster University of Applied Sciences,
Stegerwaldstrasse 39, 48565 Steinfurt, Germany}
\affiliation{$^2$International Institute of Physics (IIP)
Federal University of Rio Grande do Norte
Av. Odilon Gomes de Lima 1722, 59078-400 Natal, Brazil
}
\affiliation{$^{3}$ Max-Planck-Institute for the Physics of Complex Systems, 01187 Dresden, Germany
}

\begin{abstract}
Based on the quantum kinetic equations for systems with SU(2) structure,
regularization-free density and pseudospin currents are calculated in graphene 
realized as the
infinite mass-limit of electrons with quadratic dispersion and a proper
spin-orbit coupling. Correspondingly the currents possess no quasiparticle
part but only anomalous parts. The intraband and interband conductivities are
discussed with respect to magnetic fields and magnetic domain puddles. It is
found that the magnetic field and meanfield of domains can be represented by
an effective Zeeman field. For large Zeeman fields the dynamical
conductivities become independent of the density and are universal in this
sense. The different limits of vanishing density, relaxation, frequency, and
Zeeman field are not interchangeable. The optical conductivity agrees well
with the experimental values using screened impurity scattering and an
effective Zeeman field. The universal value of Hall conductivity is shown to
be modified due to the Zeeman field. The pseudospin current reveals an anomaly since a quasiparticle part appears though it vanishes for particle currents. The density and pseudospin response functions to an external electric field are calculated and the dielectric function is discussed with respect to collective excitations. A frequency and wave-vector range is identified where the dielectric function changes sign and the repulsive Coulomb potential becomes effectively attractive allowing Cooper pairing. 
\end{abstract}
\maketitle

\section{Introduction}

Graphene has been the topic of tremendous theoretical activity with many
complete reviews published \cite{NGPNG09,AABZC10,SAHR11}. The question is 
what one
might find new besides these many excellent investigations. In order to provide a systematic approach allowing successively better approximation, one aim is to
describe all transport properties and excitation properties with the help of
a single theoretical method which will be here the quantum kinetic theory. Another goal is to
explore new branches of excitations which might not be treated yet. Here it will be
presented a quantum kinetic approach to these transport properties including magnetic fields and meanfields due to magnetic domains. The range of
parameters is explored where the effective Coulomb interaction between the electrons in
graphene might change the sign. This allows to pair the
electrons as a necessary condition for superconductivity.

In graphene the chiral nature of the charge carriers leads to a minimal finite conductivity even with vanishing density of scatterers. If there are no charge carriers the field has to create first electron-hole pairs before they can be accelerated. Since the absolute value of the velocity is fixed, only the direction can change which provides an anomaly transport \cite{KLR10}. This remarkable feature of dissipation in an even ideal crystal is reached in various limiting procedures.
The static limit and vanishing relaxation rate are not interchangeable as it was noted \cite{Lud94}. If one first take the static limit and than the collision free limit one obtains
\be
\sigma_1=g{e^2\over 2 \pi^2 \hbar}
\ee
while the opposite order provides
\be
\sigma_2=g{e^2\over 16 \hbar}
\label{s2}
\ee
which agrees within $1-2\%$ with experiments \cite{Nov05,Nair08,KHCM08} taking into account the valley degeneracy $g=4$. Sample-dependent prefactors \cite{Tan07} have been attributed to inhomogeneous charge distributions around the Dirac point \cite{Hwang07}. Twice $\sigma_2$ is obtained if one takes the limit $\omega=\hbar/\tau$ \cite{Zie06,Z07}.

This different limiting values have provoked many theoretical investigations \cite{SAHR11}. Both results have been obtained by Kubo formula approaches depending on the order of limits performed.  In \cite{Cs07} the result $\sigma_2$ was obtained with the factor of 2 for bilayer graphene and the result for multilayer graphene was given in \cite{KBT13}. In \cite{TS08} the minimal conductivity was analyzed with respect to long and short-range scatterers and differing measurements are explained by the dominance of either of these scatterers. The density matrix approach \cite{CW08} discusses the parallels between steady states reached in graphene and the precession motion in spin-orbit coupled systems. The surface state of a topological insulator with the effective Hamiltonian $D p^2+\lambda \V \sigma \cdot \V p$ has been treated in \cite{CHSTS10} where
the limit $D \to 0$ leads to the correct result. It is argued that the transition between both regimes is due to low and high density limits. We will use the limit of infinite mass to extract the specific results for graphene from results of the general transport theory for SU(2) systems \cite{M15}.

The tight binding approach in electric fields \cite{KLR10} provides essentially the correct expression $\sigma_2$ and discusses two theoretical classes leading to the wrong result $\sigma_1$. Essentially there is no small parameter as the usual Ioffe rule $\hbar/\epsilon_F\tau$ and diagrammatic approaches miss probably important diagrams. The second class of treatments leading to $\sigma_1$ relies on the Landauer approach and takes the limiting value of finite width and length of the ribbon sample. This is performed by counting evanescent modes \cite{Lou15}. There it was claimed that the appearance of the minimal conductivity is due to the topological property of the Berry phase and nature of evanescent waves and is not a signal of Zitterbewegung and chiral symmetry as suggested in \cite{K06,WHX12}

The influence of impurity and phonon scattering has been investigated in high electric fields \cite{BLLW09} within a balance equation approach. It is suggested that the electron-electron interaction or strains in the sample causes a symmetry breaking which was modeled by a complex parameter \cite{DROC12} leading to $\sigma_1$.

The expansion around the correct value $\sigma_2(1+C\alpha_V)$ in terms of
fine structure constant of graphene $\alpha_V=e^2/(4 \pi \hbar v)\approx 2.2$
had raised a controversy. The two loop interaction correction was calculated
in two ways, via density-density and current-current correlation functions
with the help of dimensional regularization \cite{TK14}. In \cite{RLM13} it
was claimed to have found the reason for the discrepancy in the first-order
expansion by calculating a tight binding approach before reduction to
mass-less Dirac particles. The different results are dependent on the used
regularization schemes. The sharp momentum cut-off used by \cite{HJVO08}
leading to $C=25/12-\pi/2$ had been criticized by Mishenko \cite{M08}
obtaining $C=19/12-\pi/2$ due to soft cut-off. This has been supported by
\cite{sh09} which states the importance of charge conservation at all stages
of approximation and completion of Ward identity. The reason for the
discrepancy with the sharp cut-off has been traced in \cite{GCCNS13} to be due
breaking of transversality of the polarization tensor which is a consequence
of the spatial O(2) symmetry. The soft cut-off result has been reproduced
there with a regularization-independent framework discussing Coulomb
corrections \cite{JVH10}.
 A first-order interaction correction to the polarization function and dielectric function was treated in \cite{SF12} and compared with other renormalization approaches. We will present a scheme which is free of any regularization.

Recently there has been tremendous effort to modify the graphene sheets in
order to create a gap which is necessary for optical applications
\cite{NGPNG09,F15}. Nanoribbons \cite{KG08,RDBTLBS14}, twisted bilayers
\cite{CW11}, disordered graphene \cite{PGN06},
or nanofluids \cite{Wang2015141,LSGG15} are some possibilities. In this
context the suggestion was made that on the surface charge puddles appear \cite{ZBGZC09,LHS11,BM15}. Motivated by this observation magnetic impurities are considered here to form magnetic domains. We model such domains assuming that this local magnetic impurities are randomly distributed on different
sites within an angle $\theta_l$ from the $\V e_z$ direction. The directional
average \cite{CH98} leads then to 
\be
\overline{\sum\limits_p f \V V}=|V|{\sin \theta_l\over \theta_l} \V e_z n(\V q,t)=\V
V(q) n(\V q,t)
\nonumber\\
\overline{\sum\limits_p \V g \cdot \V V}=|V|{\sin \theta_l\over \theta_l} \V e_z \cdot 
\V s(\V q,t)=\V V(q) \cdot \V s(\V q,t)
\label{direct}
\ee 
where the Fourier transformed, $\V r\to \V q$, time-dependent density and polarization density are given by
\be 
\sum_p f(\V p,\V q,t)=n(\V q,t),\qquad
\sum_p \V g(\V p,\V q,t)=\V s(\V q,t)
\label{quasin}
\ee
and $\sum_p=\int d^Dp/(2\pi\hbar)^D$ for $D$ dimensions. The magnetization density becomes
$\V M(\V r,t)=g  \mu_B \V s(\V r,t)$.

The angle $\theta_l$ in (\ref{direct}) allows us to describe different models. A completely
random local magnetic field $\theta_l=\pi$ is used for magnetic impurities in
a paramagnetic spacer layer and in a ferromagnetic layer one uses
$\theta_l=\pi/4$. The latter one describes the randomly distributed
orientation against the host magnet \cite{CH98}. Therefore the same impurity
potential appears as magnetic impurities. We will see that it adds to the
Zeeman field and we will call it effective Zeeman field during this paper.

The outline of the paper is as follows. First we sketch shortly the quantum kinetic approach including electric and magnetic fields and how the known results for particles with quadratic dispersion can be translated into the specific linear dispersion of graphene.
Then we discuss the currents in section III. From the linearization of the kinetic equation we get the response function for the coupled density and pseudospin response with respect to an external electric field. In section V we discuss the intraband and interband conductivity in detail with various limiting ranges. There we generalize known results now including magnetic fields, magnetic domains and meanfields. The comparison of the longitudinal optical conductivity is presented and the Hall conductivity is discussed. Section VI contains the pseudospin conductivity and shows a subtlety in using the infinite-mass limit in that contrary to the particle current the normal quasiparticle pseudospin current does not vanish. Finally section VII explores the region where the effectively screened Coulomb repulsion between electrons is changing sign opening the possibility to form Cooper pairs. The summary concludes the paper.

\section{Kinetic theory}

We consider an effective Hamiltonian possessing a Pauli structure
\be
\hat H_{\rm eff}=H+\V \sigma\cdot \V \Sigma
\ee 
with the Pauli matrices $\V\sigma$ and the scalar Hamiltonian
\be
H={p^2\over 2 m}+\Sigma_0(\V p,\V r,t)+e \Phi(\V r, t)
\ee
consisting of 
a quadratic dispersion of particles and scalar meanfield selfenergy $\Sigma_0$, the scalar e.m. potential $\Phi$ and the vector potential absorbed in the canonical momentum $\V p=\V k-e \V A$. Any spin-orbit coupling $\V b$ and magnetic field $\V B$ can be written compactly as \cite{M15}
\be
\V {\Sigma}={\V \Sigma}^{H}(\V p,\V r,t)+\V b(\V p)+\mu_B \V B.
\label{sig}
\ee
The Hartree meanfield selfenergies read \cite{M15}
\be
\Sigma_0(p,q,t)&=&
n(q,t) V_0(q)+\V s(q,t)\cdot \V V(q)
\nonumber\\
\V \Sigma^H(p,q,t)&=&
\V s(q,t) V_0(q)+n(q,t) \V V(q)
\label{mf1}
\ee
where $\V V$ describes the averaged magnetized domains or magnetic impurities (\ref{direct}). Together with (\ref{quasin}) and the kinetic equation for $f$ and $\V g$ they form a selfconsistent equation system which consequences \cite{M15} are not the topic here. We include the meanfield to see where it appears and need it to derive the response function later.

In graphene we can consider the two sublattices described by the value of the z-component of the Pauli matrix \cite{Kane:2005,SSS12}. Since the two K-points are not coupled \cite{SSS12}  we account for by the degeneracy $g=2$ and an additional factor $2$ for spin degeneracy. For single-layer graphene the above form of spin-orbit coupling can be considered as pseudo-spin representing the linear dispersion for the kinetic energy 
$\V b=v (p_x,p_y,0)$ and the limit of $m\to\infty$. We will use this procedure to see how the results for spin-orbit coupled systems of \cite{M15,M15a} translate into graphene. The (pseudo)spin-Hall effect in graphene itself is also investigated \cite{Kane:2005} reporting the anomalous Hall effect in single-layer and bilayer graphene \cite{PhysRevB.82.161414,PhysRevB.83.155447} and which is treated like a spin-orbit coupled system \cite{Tk09} too.

The Wigner distribution function consists now of a scalar and a vector part 
\ba
\hat \rho(\V p,\V r,t)=\!{\rm Tr} \hat \rho_S \Psi^+\Psi =f \!+\!\V \sigma \cdot \V g
=\!\begin{pmatrix}
f\!+\!g_z &
g_x\!-\!i g_y
\cr 
g_x\!+\!i g_y 
&f\!-\!g_z
\end{pmatrix}
\label{rhofg}
\end{align}
with (\ref{quasin}) and the spinor creation operator $\Psi^+=(\Psi^+_\uparrow,\Psi^+_\downarrow)$ with the trace over the nonequilibrium  statistical operator $\hat \rho_S$. 

The quasi-classical kinetic equations for this mean-field Hamiltonian consist of two coupled equations \cite{M15}
\be 
D_t f+\V A \cdot \V g &=&0\nonumber\\
D_t \V g+\V A f
&=&2 (\V \Sigma\times \V g)
\label{kinet}
\ee
where $D_t=(\p t+\V {\cal F}\V {\p p}+\V v\V {\p r})$ describes the drift and force of the scalar and vector part with the velocity 
\be
v={p\over m_e}+\p p \Sigma_0
\label{vel}
\ee
and the effective Lorentz force
\be
\V {\cal F}=(e \V E +e \V v \times  \V B - \V {\p r}
\Sigma_0).
\label{lor}
\ee
The coupling between spinor parts is given by the vector drift
\be
A_i=\V \partial_p \Sigma_i\cdot \V \partial_r-\V \partial_r\Sigma_i\cdot \V \partial_p+e(\V \partial_p\Sigma_i \times \V B)\cdot \V \partial_p.
\label{A}
\ee
Remember that we have subsumed in the vector selfenergy (\ref{sig})
the magnetic impurity meanfield, the spin-orbit coupling vector, and the Zeeman term.

The term (\ref{A}) in the second parts on the left sides of (\ref{kinet}) represents the coupling between the spin parts of the Wigner distribution. The vector part contains additionally the spin-rotation term on the right hand side. 
One has to consider additionally collision integrals which have been derived in \cite{SMEH06, RGSD06, GSRS10}. In the simplest way we will add a relaxation time with conserving Mermin's correction \cite{MER70,Ms12}. 

\begin{figure}
\includegraphics[width=8cm]{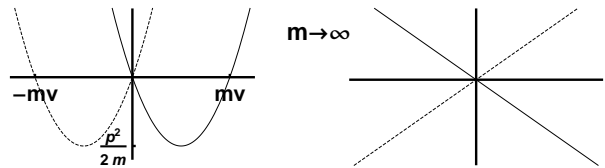}
\caption{\label{limit}
Realization of linear Dirac dispersion of graphene by the infinite-mass limit of spin-orbit coupled system with quadratic dispersion.}
\end{figure}

The stationary solution of (\ref{kinet}) has the structure \cite{M15}
\ba
\hat \rho({\hat \varepsilon})=\sum\limits_{\pm}\hat P_\pm f_\pm
&=&{f_+\!+\!f_-\over 2}\!+\!\V \sigma\cdot \V e \,\,{  f_+\!-\!f_-\over 2}
\equiv f+\V \sigma \cdot \V g
\label{solF}
\end{align}
with $f_\pm=f_0(\epsilon_p(\V r)\pm|\V \Sigma(\V p,\V r)|)$ and in equilibrium $f_0$ is the Fermi-Dirac distribution. The self-consistent effective spin-polarization direction \cite{M15},
\be
\V e ={\V \Sigma\over |\V \Sigma|},
\ee
 is given by the vector part of the selfenergy (\ref{mf1}), the magnetic field and the $\V b$-vector combined into an effective Zeeman field (\ref{sig}).
We obtain obviously a splitting of quasiparticle energies due to spin-orbit coupling 
\be
\epsilon_\pm={p^2\over 2 m}+\Sigma_0 \pm |\V \Sigma|\rightarrow \pm v p
\ee
which takes the form of single-layer graphene for $m\to \infty$ and vanishing magnetic and mean fields $\V \Sigma=\V b=v\V p$.
This supports the idea to represent the single-layer graphene simply by the limit of infinite mass as illustrated in figure \ref{limit} and used in \cite{CHSTS10}. Please note that during this limit the bounded dispersion from below turns into an unbounded Luttinger-type of dispersion. Therefore this limit changes the structure of equations in an nontrivial way. In fact we will see that various limits cannot be interchanged with this infinite-mass limit.

\section{Currents}
Due to the spin-orbit coupling (\ref{sig}) the current possesses a normal and anomaly part. Using $[\V b(\V p),x_j]=-i \hbar \partial_{p_j} \V b(\V p)$ from elementary quantum mechanics we have 
\be
\hat v_j=\frac i \hbar [\hat H,\hat {x}_j]
=\p {p_j} \epsilon+ \partial_{p_j} \V b \cdot \V \sigma
\label{18}
\ee
if the single particle Hamiltonian is given by the quasiparticle energy $\epsilon(p)$.
Together with the Wigner function (\ref{rhofg}) one has
\ba
\hat \rho \hat v_j=f \p {p_j} \epsilon+\V g \cdot  \partial_{p_j} \V b +\V \sigma\cdot (\p {p_j} \epsilon\V g+f \partial_{p_j} \V b +i \partial_{p_j} \V b \times \V g)
\end{align}
and the particle and pseudospin current densities read
\ba
\hat j_j&= \sum\limits_p [\hat \rho, v_j]_+
\nonumber\\
&=2 \sum\limits_p\left [
f \p {p_j} \epsilon\!+\!\V g \cdot \partial_{p_j} \V b \!+\!\V \sigma\cdot (\p {p_j} \epsilon\V g\!+\!f \partial_{p_j} \V b )
\right  ]\nonumber\\
&=j_j+\V \sigma\cdot \V S_j.
\label{current}
\end{align}
The scalar part describes the particle current $\V j=\V j^n+\V j^a$ consisting of a normal and anomaly current and the vector part describes the pseudospin current $S_{ij}$ not to be confused with the polarization $\V s$. 

In the limit of infinite mass, $\epsilon_p\to 0$ and therefore $\p {p_j} \epsilon\to 0$ we obtain that graphene can only possess an anomalous particle current since the normal one would be of Drude type vanishing for $m\to\infty$. The normal pseudospin current, however, possesses a finite $m\to\infty$ limit which is unexpected. This we will treat in section \ref{spincurrent}. 

Let us first consider the particle current. 
In \cite{M15} the linearized solution of (\ref{kinet}) has been derived with respect to a time-varying electric field. The Fourier transform $t\to\omega$ reads in the long-wavelength limit [$E\partial_p=\V E_\omega\cdot \V \partial_p$]
\be
\delta \V g(\omega,p)&=& 
{i \omega \over 4 |\Sigma|^2-\omega^2} e E\partial_p\V g
\nonumber\\&&
-2  {1  \over 4 |\Sigma|^2-\omega^2} \V \Sigma \times eE\partial_p \V g
\nonumber\\&&
-4 i{1\over \omega (4 |\Sigma|^2-\omega^2)}\V \Sigma (\V \Sigma \cdot eE\partial_p \V g).
\label{deltarho}
\ee 
From the three terms in (\ref{deltarho}) we get the symmetric and asymmetric conductivity \cite{M15}. One has according to (\ref{current})
\ba
\delta j_\alpha^a=\sum\limits_p \partial_{p_\alpha} \V \Sigma\cdot \delta \V g
=\sum\limits_p \Sigma \partial_{p_\alpha} \V e \cdot \delta \V g+\sum\limits_p(\V e \cdot \partial_{p_\alpha} \V \Sigma)\V e\cdot \delta \V g. 
\label{dj0}
\end{align}
Please note that $\p p \V \Sigma=\p p\V b$.
Writing $j_\alpha=\sigma_{\alpha \beta } E_\beta$ we obtain from the first part of (\ref{dj0}) using (\ref{deltarho}) the interband and the anomalous Hall conductivity
\be
\sigma_{\alpha\beta}^{\rm inter}&=&
2 {e^2}\sum\limits_p {g  \over
  1-{\omega^ 2\over 4|\Sigma|^2}}\, 
{i\omega\over
    2|\Sigma|}\partial_{p_\alpha} \V e\cdot  \partial_{p_\beta} \V e
\nonumber\\
\sigma_{\alpha\beta}^{\rm Hall}&=&
2 {e^2}\sum\limits_p {g  \over
  1-{\omega^ 2\over 4|\Sigma|^2}}\,
\V e\cdot (\partial_{p_\alpha} \V e\times \partial_{p_\beta} \V e)
\label{cond1}
\ee
where the first part of (\ref{deltarho}) leads to the interband and the second part of (\ref{deltarho}) to the Hall conductivity. 

The second part of (\ref{dj0}) with (\ref{deltarho}) leads to the intraband conductivity
\be
\sigma_{\alpha\beta}^{\rm intra}=
i 2 e^2\sum\limits_p \partial_{p_\alpha}\partial_{p_\beta} \Sigma {g\over \omega}
\label{cond2}
\ee
where the first and the third part of (\ref{deltarho}) contributes.
This intraband contribution had been neglected in \cite{M15}.

For graphene we have $\partial_{p_\alpha} \Sigma_i=v \delta_{\alpha,i}$ and the anomalous current (\ref{dj0}) reads together with the linear response (\ref{deltarho}) to any time-dependent electric field
\ba
&\delta \V j^a(t)=2 e v \sum\limits_p \!\delta \V g=-2 e v \sum\limits_p\!\int\limits_0^t \! d\bar t {\rm e}^{-{{\bar t}\over \tau}}\!\biggl \{ \cos{(2 v p \bar t)}E_{t\!-\!\bar t} \partial_p \V g
\nonumber\\&
\!+\! \sin{(2 v p \bar t)}\V p_0\times E_{t\!-\!\bar t}\partial_p \V g
\!+\![1\!-\!\cos{(2 v p \bar t)}]
\left [
\V p_0\cdot E_{t\!-\!\bar t}\partial_p \V g
\right ]
\V p_0 \biggr \}
\label{jtime}
\end{align}
with $\V p_0={\V p/ |\V p|}$.
Each term from (\ref{deltarho}) corresponds therefore to a specific precession motion analogously to the one seen in the conductivity of a charge in crossed electric and magnetic fields. If we replace $\omega_c=e B/m\leftrightarrow 2 |\Sigma|= 2 v p$ in (\ref{jtime}) one has
\ba
&e\delta \V j(t)=\sigma_0 \int\limits_0^t {d\bar t\over \tau} {\rm e}^{-{\bar t\over \tau}}\left \{ \cos{(\omega_c \bar t)}\V E(t-\bar t)
\right .\nonumber\\&\left .
\!+\! \sin{(\omega_c \bar t)}\V E(t\!-\!\bar t)\times \V B_0
\!+\![1\!-\!\cos{(\omega_c \bar t)}][\V E(t\!-\!\bar t)\cdot \V B_0]\V B_0\right \}
\label{jtimec}
\end{align}
which is just the solution of the Newton equation of motion
\be
m_e \dot {\V v}=e (\V v\times \V B)+e \V E-m_e {\V v\over \tau}.
\label{start}
\ee
It illustrates the threefold orbiting of the electrons with cyclotron frequency: (i) in the direction of the electric and (ii) magnetic field, and (iii) in the direction perpendicular to the magnetic and electric field.  

One can find these 3 terms also directly solving the Heisenberg equation for the spinor creation operators from the free graphene Hamiltonian 
\be
v\sum\limits_p\Psi_p^+\V \sigma\cdot \V p \Psi_p
\ee
which reads \cite{K06}
\be
\Psi_t=\left [\cos (v p t)-i {\V \Sigma\cdot \V p\over p} \sin (v p t)\right]\Psi_p.
\label{psit}
\ee
The charge current operator (\ref{18}) is calculated directly \cite{K06} with (\ref{psit})
\be
\hat {\V j}&=&e v \sum\limits_p \Psi_t^+\V \sigma \Psi_t
\nonumber\\
&=&e v \sum\limits_p \Psi_p^+\left [ \cos (2 v p t) \V \sigma+(1-\cos(2 v p t)) (\V \sigma \cdot \V p) \V p 
\right .\nonumber\\&&\left . 
+\sin(2 v p t) (\V p\times \V \sigma)\right ]\Psi_p.
\label{jop}
\ee
We use now the ensemble average 
\be
{\rm Tr} \hat \rho_S \Psi_\alpha^+\V\sigma_{\alpha \beta}\Psi_\beta=\V \sigma_{\alpha \beta} (f_0 \delta _{\alpha \beta}+\V \sigma_{\beta\alpha} \times \V g_0)=2 \V g_0
\ee
with the Wigner function (\ref{rhofg}) we obtain exactly (\ref{jtime}) with $\V g_0=eE\partial_p \V g$.
There is no reason to call the last term in (\ref{jop}) an expression of Zitterbewegung as claimed in \cite{K06}.

\section{Response functions}

If one linearizes the kinetic equation (\ref{kinet}) with respect to an external electric potential $\Phi$ one obtains the equation system for density and pseudospin responses \cite{M15a}
\ba
\left (1-{i\over \bar \omega \tau}\right ) \delta n &=\Pi_0 (\delta \Sigma+\Phi)+\V \Pi \cdot \delta \V \Sigma+\sum\limits_p {q\partial_p \V \Sigma\cdot \delta \V g\over \tilde \omega}\nonumber\\
\delta \V s&=\V \Pi_3 (\delta \Sigma \!+\!\Phi)\!+\!\V \Pi_2\times \delta \V \Sigma\!+\!\Pi_0 \delta \V \Sigma\!+\!\overleftrightarrow \Pi\cdot \V \delta \Sigma
\label{res}
\end{align}
for any spin-orbit coupling where the different polarization functions are given in \cite{M15a}, $\tilde \omega =\omega -\V p\cdot \V q/m +i/\tau$, $\bar \omega=\omega+i/\tau$, and the intrinsic mean field (\ref{mf1}) variations read
\be
\delta \V \Sigma &=&\V V \delta n+V_0 \delta \V s\nonumber\\
\delta \Sigma&=&\V V\cdot \delta \V s+V_0 \delta n.
\ee
Here we represent any interaction of electrons with scalar impurities or among themselves by the potential $V_0$ and with magnetic impurities by the potential $\V V$. The latter allows to include averaged magnetization domains. The term on the left hand side of (\ref{res}) in front of the density variation is a result of Mermin's correction \cite{MER70,Ms12} and crucial for conserving relaxation time approximation.

For graphene in the infinite-mass limit the response functions take the form $\Pi_0=\V \Pi=0$ and 
\be
\V \Pi_3&=&{2 i \Sigma_n\lambda \over v \bar \omega} {\cal E} \V e_z\times \V q- {4 \lambda^2\over \bar \omega} {\cal C} \V q\nonumber\\
\V \Pi_2&=&-{2 i \Sigma_n \over v \bar \omega} {\cal E} \V e_z- {i \lambda\over \bar \omega} {\cal F} \V q\nonumber\\
\overleftrightarrow \Pi\cdot \V \delta \Sigma&=&-{4\lambda \over \bar \omega} \left ({\cal C} \delta \V \Sigma_\perp+\tilde {\cal C} \delta \Sigma_z\V e_z\right )
\nonumber\\&&
+{2\lambda^2 \Sigma_n\over v \bar \omega} {\cal F} \left ( \delta \Sigma_z\V q+(\V q\cdot \delta \V \Sigma) \V e_z\right ).
\label{pol}
\ee
We used $\lambda=v/\bar \omega\hbar$, $|\V \Sigma|=v \Gamma$ with $\Gamma^2=p^2+(\Sigma_n/v)^2$, and the effective Zeeman field in $z$-direction 
\be
\Sigma_n=|n \V V+\V s V_0+\mu_B \V B|.
\label{147}
\ee
The direction of pseudospin is $\V e=(p_x,p_y,{\Sigma_n\over v})/\Gamma$, and the used integrals read performing angular integrations
\be
{\cal A}&=&\sum\limits_p {g\over \Gamma} \left (1-{p^2\over 2 \Gamma^2}\right
)=-\sum\limits_p {p^2\over 2 \Gamma^2} f'=-{\cal D},\nonumber\\
{\cal B}&=&\sum\limits_p {g\over \Gamma} {1-{p^2\over 2 \Gamma^2}\over 1-4 \lambda^2\Gamma^2}={\cal A}+4 \lambda^2 {\cal C},\nonumber\\
{\cal C}&=&\sum\limits_p {g \Gamma \left (1-{p^2\over 2 \Gamma^2}\right )\over 1-4 \lambda^2\Gamma^2},
\quad \tilde {\cal C}=\sum\limits_p {g \Gamma \left (1-{B^2\over\Gamma^2}\right )\over 1-4 \lambda^2\Gamma^2},
\nonumber\\
{\cal E}&=&\sum\limits_p {g\over \Gamma \left (1-4 \lambda^2\Gamma^2\right )},
\quad
{\cal F}=\sum\limits_p {p^2 g'\over \Gamma \left (1-4 \lambda^2\Gamma^2\right )}
\label{const}
\ee
with $f'=v \partial_\epsilon f$. The nontrivial identity in the first line can be shown from $\sum_p (\V q\cdot \V \partial_p) \V g=0$. The forms (\ref{const}) are convenient objects for numerical evaluation at finite temperatures. Please note that these forms are much more rich than the pure mean-field free polarizations found in \cite{KM16}.

While ${\cal A}$ and ${\cal C}$ seem to be divergent needing a cut-off due to unbounded hole states, ${\cal B}$ is convergent. However, {\cal D} is convergent as seen below which illustrates the delicate compensation of divergences in ${\cal A}$ and ${\cal C}$ due to the angular integration.
Actually, inspecting (\ref{const}) it seems that due to $f_-$ the ${\cal A}$ term is divergent for large momenta. 
Usually this has been regularized by an upper momentum cut-off representing the band width. Let us inspect how this procedure appears in the density $n$ and polarization $s$
\be
n/s=\frac 1 2 \sum\limits_p (f_+ \pm f_-)
\ee
with $f_\pm=(1+{\rm e}^{\pm v p-\mu\over T})^{-1}$. The upper integration limit of $f_-$ diverges. One interprets $1-f_-=(1+{\rm e}^{v p+\mu\over T}+1)^{-1}=f_h$ as the distribution of holes \cite{BSGTNM15,AK07} which splits the density into the baryon density $n_p-n_h$
\be
n/s=n_p\mp n_h \pm{p_\lambda^2\over 8 \pi  \hbar^2}
\ee
and a part given by the whole momentum sum with the upper momentum cut-off
$p_\lambda$. At zero temperatures we have in two dimensions
\be
n(T=0)=n_p=s={\mu^2\over 8 \pi \hbar^2 v^2}
\ee
with the chemical potential (Fermi energy) $\mu$.

This regularization seems to be applicable to the expression ${\cal A}$ as done in the literature. However, there is a delicate balance rendering this integral finite without regularization due to the identity expressing ${\cal A}$ by the obviously converging expression ${\cal D}$.
Let us inspect the structure of this term by splitting off a convergent magnetic-field-dependent part
\be
{\cal D}&=&{v\over 2}\sum\limits_p{p^2\over \Gamma^2}\partial_\epsilon f
\nonumber\\
&=&
{v\over 2}\sum\limits_p\partial_\epsilon f-{\Sigma_n^2\over 2 v}\sum\limits_p{\partial_\epsilon f\over \Gamma^2}
\nonumber\\
&=&
-{v\over 2}\partial_\mu n-{\Sigma_n^2\over 2 v}\sum\limits_p{\partial_\epsilon f\over \Gamma^2}.
\label{D1}
\ee
This shows that the divergent part is represented by the compressibility. 
 Assuming that this momentum cut-off is solely given by the band width and not dependent on the density, i.e. chemical potential, we get the first part of (\ref{D1}) analytically 
\be
-{v\over 2}\partial_\mu n&=&{1\over 4 \pi \hbar^2}\left \{ -\Sigma_n{\sinh{\Sigma_n\over T}\over \cosh{\mu\over T}+\cosh{\Sigma_n\over T}}
\right .\nonumber\\ &&
\left .+T \ln\left [ 2 (\cosh{\mu\over T}+\cosh{\Sigma_n\over T})\right ]\right \}.
\ee
The cancellation of the divergence comes here due to the angular integration.

\section{Conductivity}

\subsection{Longitudinal conductivity}
The interband and intraband (\ref {cond1}) conductivities provide both parts of the longitudinal conductivity
\be
\sigma_{xx}&=&-i e^2 2\lambda ({\cal B}+{\cal D} )
\label{con1}
\ee
where the first part is the interband and the second part the intraband contribution.  We will discuss both parts separately.

First we prove the internal consistency of the theory by deriving the conductivity from the response function. The conductivity is defined as the response with respect to the true internal electric field $E$ in contrast to the $\epsilon \epsilon_0 E$ field having external sources. This means we have to consider the response without meanfields, i.e. the polarization function $\Pi={\delta n(\Sigma=0)/ \Phi}$. Then (\ref{res}) reduces to
\be
\left (1-{i\over \bar \omega \tau}\right ) \delta n&=& \lambda \V q \cdot \V \delta s
\nonumber\\
\delta \V s&=& \V \Pi_3 \Phi
\label{res0}
\ee
which yields with  (\ref{pol}) exactly the current (\ref{jtime}). As a cross check we use for $V_0$ the Coulomb potential. From the Maxwell equation 
the longitudinal conductivity is expressed as 
\be
\sigma_{xx}=-\epsilon_0\omega V_0 {\rm Im} \Pi
\ee
and from (\ref{res0})
\ba
\sigma_{xx}=-\epsilon_0\omega V_0{\rm Im} \left ( -{4 \lambda^3 {\cal C}
    q^2\over \bar \omega-{i\over \tau}}\right )=4 e^2 l_D {\cal C} v^3{\rm Im} (i \tau_\omega^3)
\end{align}
with 
\be
\tau_\omega= {\tau\over (1-i \omega \tau)}
\label{tw}
\ee 
and $l_D=1$ for three dimensions and $l_D=q/2\hbar$ for quasi-two dimensions \cite{S16}.
For the static limit one gets exactly the result from current formula (\ref{con1}) since ${\cal B}+{\cal D}={\cal B}-{\cal A}=4 \lambda^2 {\cal C}$. This illustrates ones more the necessity to include Mermin's corrections \cite{MER70,Ms12} represented by the left side subtraction in (\ref{res}) or (\ref{res0}).

\subsubsection{Interband conductivity}

The zero-temperature limit of the interband part considered in \cite{M15} reads
for particles with quadratic dispersion
\ba
\sigma^{B}_{xx}&=-i e 2\lambda {\cal B}={e^2\over 8 \pi \hbar} \left \{
{4 \epsilon_v \Sigma_n^2 \tau_\omega/\hbar\over 2 \epsilon_v \mu+\Sigma_n^2}
\right .\nonumber\\ &+\left .
(1-{4 \Sigma_n^2 \tau_\omega^2\over \hbar^2})
\arctan{\left [
{4 \epsilon_v \tau_\omega \hbar\over \hbar^2+4 (2 \epsilon_v \mu+\Sigma_n^2)\tau_\omega^2}
\right ]}
\right \}
\label{148}
\end{align}
with $\epsilon_v=m v^2/\hbar$. This conductivity represents a contribution in the direction of the applied electric field and is caused by collisional correlations.
This dynamical result is different from the pseudospin accumulation found in \cite{BK06} basically by the $arctan$ term and therefore no sharp resonance feature. Expanding, however, in small spin-orbit coupling
\ba
\sigma^{B}_{xx}&={e^2\over \pi \hbar} \,{\epsilon_v \tau_\omega\over1+4 \Sigma_n^2 \tau_\omega^2}+o(\epsilon_v^2)
\end{align}
shows that the static limit agrees with \cite{BK06,IBM03}. Please note that if one sets $\Sigma_n\to 0$ before expanding a factor 1/2 appears which illustrates the symmetry breaking by the effective Zeeman term.

\begin{table}
\begin{tabular}{|c|c|c|c|c||c|}
\hline
\multicolumn{5}{|c||}{order of limits}& $\sigma_{xx}^B=\zeta {e^2\over 16 \hbar}$\\
\hline
5.&4.&3.&2.&1.&$\zeta$
\cr
\hline
\hline
$\mu\to 0$&$\omega\to 0$ &$\Sigma\to 0$ &$m\to\infty$ &$\tau\to\infty$& $-1$
\cr
\hline
$\mu\to 0$&$\tau\to\infty$ &$\Sigma\to 0$ &$m\to\infty$ &$\omega\to 0$& $0$
\cr
\hline
$\tau\to\infty$&$\mu\to 0$ &$\Sigma\to 0$ &$m\to\infty$ &$\omega\to 0$& $1$
\cr
\hline
$\mu\to 0$&$\Sigma\to 0$ &$\tau\to\infty$ &$m\to\infty$ &$\omega\to 0$& $1$
\cr
\hline
$\mu\to 0$&$\tau\to\infty$ &$m\to\infty$ &$\Sigma\to 0$ &$\omega\to 0$& $1$
\cr
\hline
$\mu\to 0$&$\tau\to\infty$ &$\omega\to 0$ &$m\to\infty$ &$\Sigma\to 0$& $1$
\cr
\hline
$\mu\to 0$&$\Sigma\to 0$ &$m\to\infty$ &$\tau\to\infty$ &$\omega\to 0$& $0$
\cr
\hline
$\mu\to 0$&$m\to\infty$ &$\tau\to\infty$ &$\Sigma\to 0$ &$\omega\to 0$& $0$
\cr
\hline
$\mu\to 0$&$m\to\infty$ &$\Sigma\to 0$ &$\omega\to 0$ &$\tau\to\infty$& $0$
\cr
\hline
$\mu\to 0$&$\Sigma\to 0$ &$\omega\to 0$ &$m\to\infty$ &$\tau\to\infty$& $0$
\cr
\hline
\end{tabular}
\caption{\label{tab2} results for different orders of limits.}
\end{table}

To translate (\ref{148}) into the formula for graphene we perform the limit of infinite mass or $\epsilon_v\to \infty$.
The order of limits becomes now essential as illustrated in table~\ref{tab2} including the vanishing Zeemann field ${\Sigma_n}\to 0$, the static limit, the infinite-mass limit, and the zero-density limit $\mu\to 0$.
If we apply the limit of vanishing friction $\tau^{-1}\to 0$ before  the infinite-mass limit we obtain $\sigma^B_{xx}=0$ and if we perform the static limit even afterward we obtain the negative result of the expected one (\ref{s2}). Since we translate the results of spin-orbit coupled systems to graphene with the help of the infinite mass limit, the latter one should be correctly performed first before any other specifications.
Performing first the infinite-mass limit we get for zero temperature
\ba
\sigma^B_{xx}={e^2\over 8 \pi \hbar}\left \{
\begin{matrix}
{2\Sigma_n^2 \tau_\omega\over \hbar \mu}+\left(1-{4\Sigma_n^2 \tau_\omega^2\over \hbar^2}\right ){\rm arccot} {2 \mu \tau_\omega\over \hbar}, & \mu>\Sigma
\cr
{2\Sigma_n \tau_\omega\over \hbar}+\left(1-{4\Sigma_n^2 \tau_\omega^2\over \hbar^2}\right ){\rm arccot} {2 \Sigma_n \tau_\omega\over \hbar}, & \mu<\Sigma
\end{matrix}
\right ..
\label{sB0}
\end{align}
Since (\ref{tw}) the limits of infinite frequency $\omega\to\infty$, vanishing scattering $\tau\to\infty$, and vanishing density $\mu \to 0$ are not interchangeable as well. In fact, if we neglect the Zeeman field, $\Sigma_n=0$, in (\ref{sB0}) we get
\be
\sigma^B_{xx}={e^2\over 16 \hbar}\left \{\begin{matrix}
-1-i {4 \mu\over \pi \hbar \omega}+o\left ({1\over \omega^2} \right )\cr
1-{4\over \pi \hbar} {\mu \tau\over \omega\tau+i}+o\left ({\mu^2 } \right )\cr
-1-i {4\mu\over \pi \hbar \omega} +o\left ({\mu}^2 \right )+o\left ({1\over \tau} \right )
\cr
0+o(\omega)+o\left ({1\over \tau} \right )
\cr
1+o\left ({1\over \tau} \right ) +o(\omega)
\end{matrix}
\right ..
\label{46}
\ee
Only the second limiting procedure leads to the right result.
One obtains even zero value if the static limit is used
after vanishing scattering which is different from interchanging both limits. 
This illustrates the care one has to take when integrating zero-temperature
values. 

\begin{figure}
\includegraphics[width=8cm]{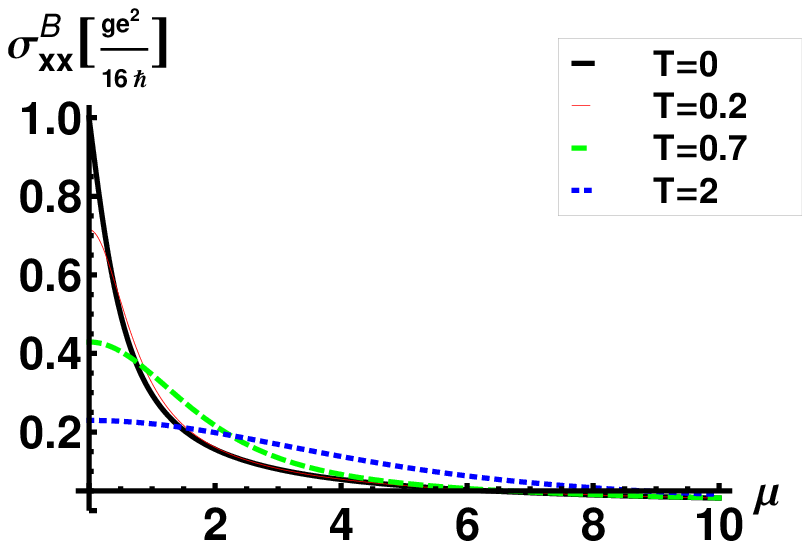}
\includegraphics[width=8cm]{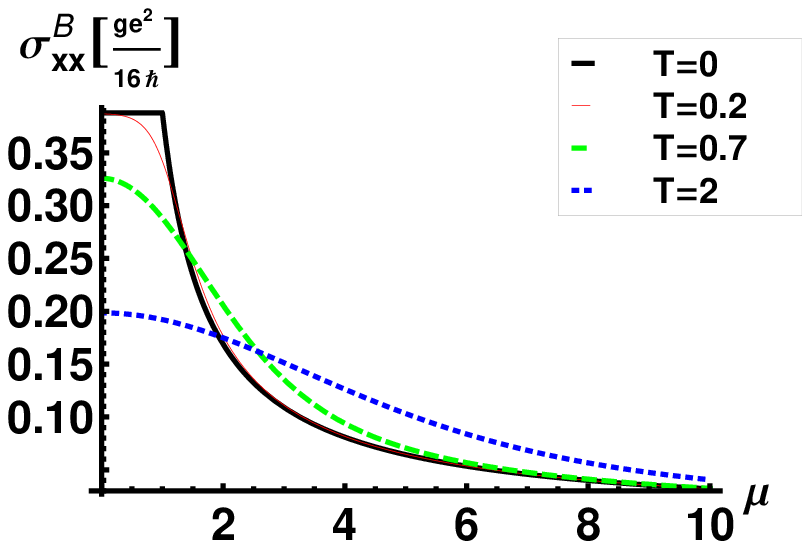}
\caption{\label{sxxB_s0_T}
(Color online) Temperature dependence (in arbitrary energy units EU) of static interband (\ref{sigB}) conductivity with $\Sigma_n=0$ (above) and $\Sigma_n=1 EU$ (below) with $\tau=\hbar/EU$ versus the chemical potential in unites EU.}
\end{figure}

The finite-temperature interband conductivity can be written with ${\cal B}$ from (\ref{const}) as
\ba
\sigma^B_{xx}={e^2 T \tau_\omega \over 4 \pi \hbar^2} \!\!\int\limits_{\Sigma/T}^\infty \!\!\!{d x} {1\!+\!\left ({\Sigma_n\over T x} \right )^2\over 1\!+\!{4 x^2  T^2\tau_\omega^2\over \hbar^2 }}\!\left ({1\over {\rm e}^{-x\!-\!\mu\over T}\!+\!1}\!-\!{1\over {\rm e}^{x\!-\!\mu\over T}\!+\!1}\!\right )
\label{sigB}
\end{align}
which is plotted in figure \ref{sxxB_s0_T}.

Please note that the frequency dependence is solely due to (\ref{tw}). We see that the interband conductivity approaches the universal value $\sigma_2$ of (\ref{s2}) in the case of no Zeeman field and $T=0$ as can be seen in (\ref{46}). For finite Zeeman fields even at zero temperature this universal value is different which rises doubts to call it universal value. As seen in figure \ref{sxxB_s0_T} from (\ref{sB0}) the finite Zeeman field introduces a density threshold below which the interband conductivity is nearly constant.

The real part of interband conductivity can be given analytically for small scattering $\tau^{-1}\to 0$ where one uses $1/(\omega+i\epsilon)=P/\omega-i \pi \delta (\omega)$. One obtains a nonzero result only for $\hbar \omega >2 \Sigma_n$ which reads 
\be
{\rm Re} \sigma_{xx}^B&=&-{e^2\over 16 \hbar} \left ( 1+{4\Sigma_n^2\over \hbar^2\omega^2} \right )
\left ({1\over {\rm e}^{{\hbar \omega\over 2 T}\!-\!\mu\over T}\!+\!1}\!-\!{1\over {\rm e}^{-{\hbar \omega\over 2 T}\!-\!\mu\over T}\!+\!1}\!\right )
\nonumber\\
&=&{e^2\over 16 \hbar} \left ( 1+{4\Sigma_n^2\over \hbar^2\omega^2} \right ){\sinh {\hbar \omega \over 2 T}\over \cosh {\mu \over T}+\cosh {\hbar \omega \over 2 T}}
\nonumber\\
&\approx&{e^2\over 16 \hbar}\left ( 1+{4\Sigma_n^2\over \hbar^2\omega^2} \right )\tanh {\hbar \omega \over 2 T},\quad \omega>2\mu\gg T.
\ee
The last expression without Zeeman field $\Sigma_n$ has been given by \cite{FV07} with a misprint of 1/4 in the argument. The second line was found in \cite{GSC09,BJZ10} within a Kubo formalism also without $\Sigma_n$. An extension of this result towards trigonal-warping corrections and bilayer graphene can be found in \cite{BZZ10}.

Due to the abbreviated dynamical relaxation time (\ref{tw}) we can now discuss the frequency dependence. For zero temperature we use from (\ref{sB0}) the case of $\mu>\Sigma_n$ since in the opposite limit one simply replaces $\mu\to\Sigma_n$ in the expression and obtains
\ba
\sigma^B_{xx}=&{e^2\over 8 \pi \hbar}\left [
{2\Sigma_n^2 i\over \mu \hbar (\omega+{i\over \tau})}
\right . \nonumber\\ & \left .
+{i\over 2} 
\left(1+{4\Sigma_n^2\over \hbar^2(\omega+{i\over \tau})^2}\right
)\ln{2\mu-\hbar (\omega+{i\over \tau})\over 2\mu+\hbar (\omega+{i\over \tau})}
\right ].
\label{sigB0}
\end{align}
If we neglect the Zeeman field $\Sigma_n\to 0$ and consider the limit of vanishing collisions $\tau \to \infty$ we obtain the result
\ba
\sigma^B_{xx}={e^2\over 16 \hbar}&\left [
\Theta(\hbar \omega-2\mu)+{i\over \pi} 
\ln \left |{\hbar \omega-2\mu\over \hbar \omega+2\mu}\right |
\right ]
\label{sBw}
\end{align}
which is exactly the result reviewed in \cite{XZLM16} with a misprint of missing $i$ and which has been first derived by \cite{AZS02} and within RPA in \cite{WSSG06}. It was extended by selfconsistent Born approximation in \cite{FKA09} and including circularly polarized light \cite{YEK15}. The Kubo formalism leads to the same expression \cite{HSZ13}. The interband conductivity (\ref{sBw}) shows a singularity at the interband transition energy $\hbar\omega=2\mu$ which is damped by collisions and finite temperatures.

Here we have extended the known expressions by the effective Zeeman field including the magnetic field, magnetized domains, and meanfields.

\subsubsection{Intraband contribution}
We have seen that the interband contribution leads to the universal low-density conductivity at zero temperature and decreases with higher densities. Now we will discuss the interband contribution which vanishes for zero temperature and low densities. In \cite{M15} we have neglected this term (\ref{cond2}) corresponding to the ${\cal D}$ intraband conductivity in (\ref{con1}). This is not justified for graphene since this term is just the intraband scattering and will lead to the linear density dependence. 

\begin{figure}
\includegraphics[width=8cm]{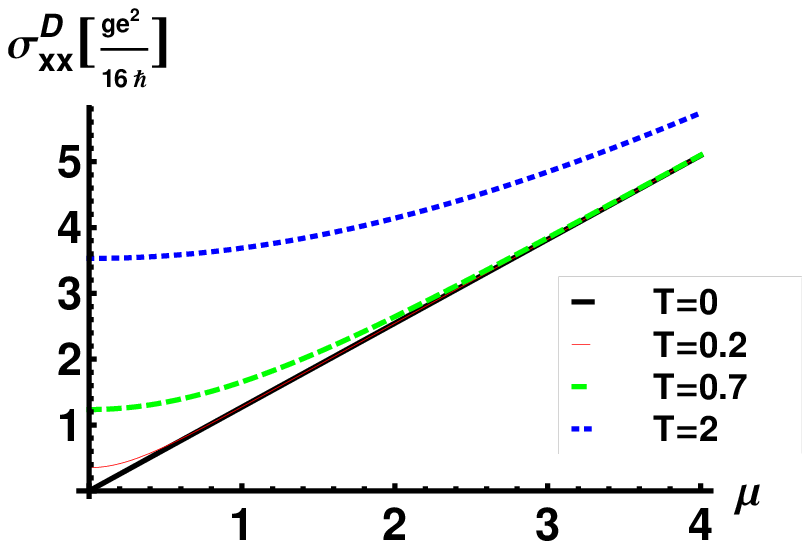}
\includegraphics[width=8cm]{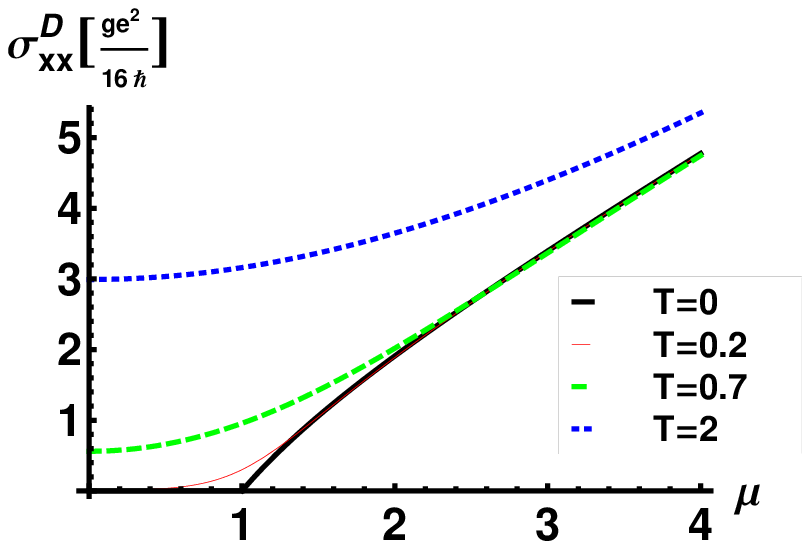}
\caption{\label{sxxD_s0_T}
(Color online) Temperature dependence (in arbitrary energy units EU) of the static intraband (\ref{sigD}) conductivity with $\Sigma_n=0$ (above) and $\Sigma_n=1 EU$ (below) and $\tau=\hbar/EU$ versus chemical potential in units EU.}
\end{figure}

The finite-temperature expression of the part (\ref{con1}) of conductivity reads explicitly
\ba
&\sigma^D_{xx}=-i e^2 2\lambda {\cal D}
\nonumber\\
&=\!{e^2 T \tau_\omega \over 4 \pi \hbar^2} \!\!\!\int\limits_{\Sigma_n\over T}^\infty \!\!\!{d x\over x} \!\!\left (\!{\Sigma_n^2\over T^2}\!-\!x^2 \!\right )
\!\!\!\left [\!
{1\over \left (\!{\rm e}^{x\!-\!\mu\over T}\!+\!1\!\right )\!\left (\!{\rm e}^{-x\!+\!\mu\over T}\!+\!1\!\right )}\!+\!(\mu\!\leftrightarrow\! -\mu)
\!\right ]
\nonumber\\
&={i\epsilon_0 \omega_p^2\over \omega+{i\over \tau}}.
\label{sigD}
\end{align}
Consequently, this intraband contribution can be written in the form of the frequency-dependent Drude conductivity with a collective frequency $\omega_p[\mu,T,\Sigma_n]$ dependent on the chemical potential, temperature and effective Zeeman field. It is illustrated in figures \ref{sxxD_s0_T} for different temperatures versus chemical potential. Except the factor $\tau_\omega\epsilon_0$ the figures~\ref{sxxD_s0_T} give therefore also the collective mode squared. One sees that the mode is excited for zero temperature only for $\mu>\Sigma_n$ which provides a threshold for densities by the effective Zeeman field. In fact, we have for zero-temperature
\be
\sigma^D_{xx}={e^2 \tau_\omega \over 4 \pi \hbar^2}\, {\mu^2-\Sigma_n^2\over \mu}\, \Theta (\mu-\Sigma_n)
\label{sigD0}
\ee
starting at chemical potentials (Fermi energies) larger than the effective Zeeman field $\Sigma_n$ seen in figures \ref{sxxD_s0_T}.

If we neglect the effective Zeeman field $\Sigma_n\to0$, the expression (\ref{sigD}) takes exactly the form of intraband contribution discussed in \cite{FV07}. Therefore we can consider the expression (\ref{sigD}) as generalization of the known result including now the magnetic field and the meanfields.

One sees that for finite temperatures the intraband conductivity is approaching a finite value for low densities. However for zero temperature this limit vanishes. Therefore the "universal" limit discussed in the literature concerns solely the intraband contribution. We can state that any finite temperature will blur this universal result since contributions start to contribute from the interband scattering.

The limit of vanishing effective Zeeman field can be given analytically as well and reads
\be
\sigma^D_{xx}(\Sigma_n=0)&=&{e^2 \tau_\omega \over 4 \pi \hbar^2}\, \left [
2 T \ln\left (1+{\rm e}^{\mu\over T}\right )-{\mu}\right ]
\nonumber\\
&=&{e^2 \tau_\omega T\over 16 \pi \hbar^2}\left (8 \ln 2+{\mu^2\over T^2} \right )+o\left ({\mu\over T}\right )^4\nonumber\\
&=&{e^2 \tau_\omega \mu\over 4 \pi \hbar^2}+o\left (T{\rm e}^{-{\mu\over T}}\right )
\label{sD}
\ee
plotted in figure \ref{sxxD_s0_T}. The expansion of the last line (\ref {sD}) agrees with Eq. 4.4 of \cite{GSh06} and the second line with Eq. 75 of \cite{GGMS02}.  

The limit of high frequencies
\be
\tau_\omega={\tau\over 1-i \omega \tau}\approx {i\over \omega}
\ee
and small densities $\mu<<T$ has been first given in the framework of 3D graphite by \cite{W47} with a corresponding additional factor $2/d$ of inverse distance between the graphite layers. As special case of (\ref{sD}) without effective Zeeman fields, the high-frequency limit coincides with the result of \cite{FV07}. 

\begin{figure}
\includegraphics[width=8cm]{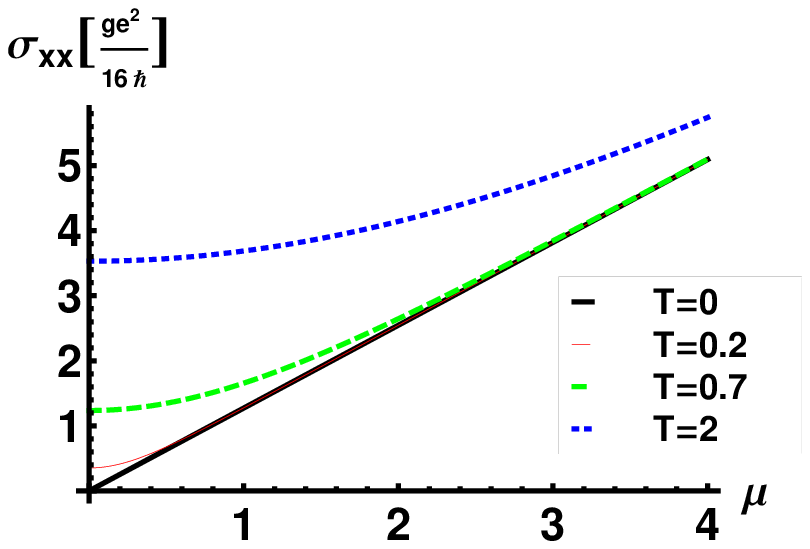}
\includegraphics[width=8cm]{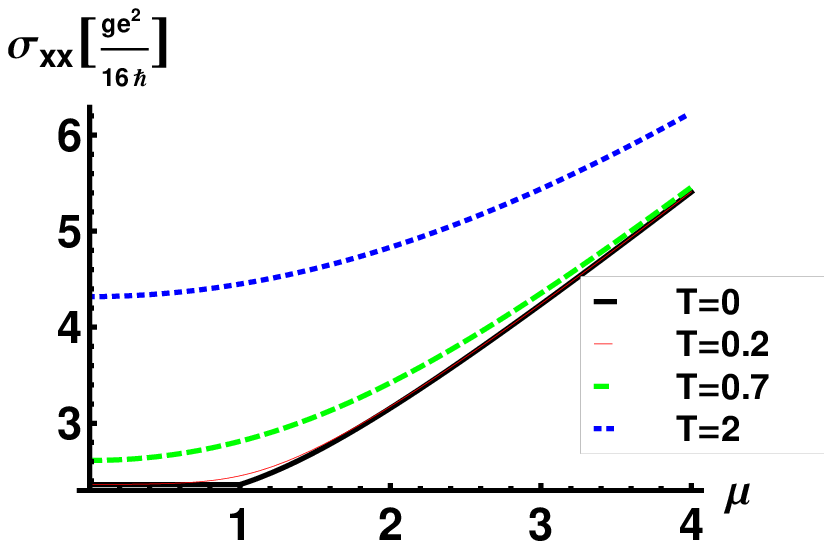}
\caption{\label{sxx_T}
(Color online) Temperature dependence (in arbitrary energy units EU) of the total (\ref{con1}) static graphene conductivity with $\Sigma_n=0$ (above) and $\Sigma_n=1 EU$ (below) and $\tau=\hbar/EU$ versus the chemical potential in units EU.}
\end{figure}

In figure \ref{sxx_T} finally we plot the complete static conductivity as a sum of (\ref{sigB}) and (\ref{sigD}). For zero temperature the universal finite value at low densities comes exclusively from the interband conductivity as discussed above. The intraband conductivity has a threshold at the effective Zeeman field. Higher temperatures mixes both components and waves the threshold of intraband conductivity.

\subsubsection{Optical conductivity: comparison with experiment}

The optical conductivity $\sigma(\omega)$ is important to know if one wants to calculate
the optical transparency \cite{SPG08}
\be
t(\omega)=\left [1+{\sigma(\omega)\over 2\epsilon_0 c}\right ]^{-2}.
\ee

\begin{figure}
\includegraphics[width=8cm]{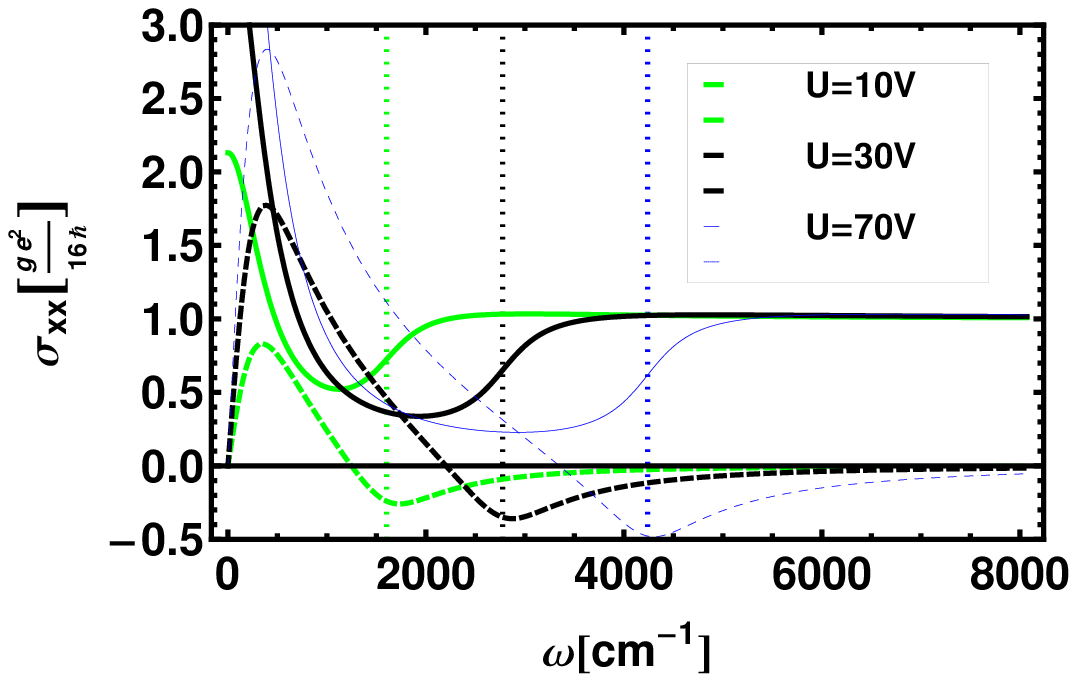}
\includegraphics[width=8cm]{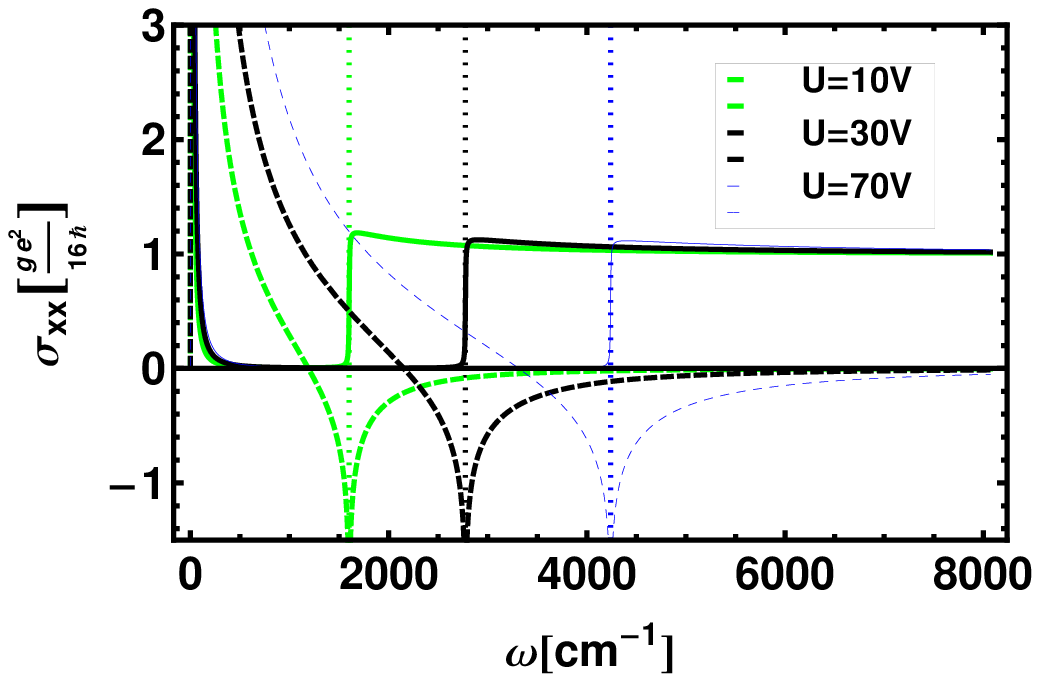}
\caption{\label{sxx_w}
(Color online) Real (solid) and imaginary (dashed) optical conductivity versus frequency for 
$\tau=20\hbar/eV$ (above) and $\tau=2000\hbar/eV$ (below). The vertical dotted lines indicate the interband threshold $\hbar \omega =2\mu$. 
The applied voltage induces a density $n={\epsilon \epsilon_0\over e d} U=7.3\times 10^{10}$cm$^{-2}$V$^{-1} U$ 
for graphene on typical SiO$_2$ substrates \cite{ZTSK05,LHJHMKS08}. The density is linked to the chemical potential 
$n={g\mu^2\over 4 \pi \hbar^2 v^2}=1010$cm$^{-2}(\mu/$eV$)^2$.
}
\end{figure}

We plot the optical conductivity at zero temperature as the sum of the interband (\ref{sigB0}) and intraband (\ref{sigD0}) conductivity and remember that the frequency dependence is given by (\ref{tw}). In figure \ref{sxx_w} we see that the imaginary part of the optical conductivity possesses a minimum at the interband threshold corresponding to a step-like behavior of the real part of the conductivity. This feature is smeared out by scattering represented by a finite relaxation time. At high frequencies the optical conductivity approaches the universal value due to interband transitions (\ref{sigB0}).

\begin{figure}
\includegraphics[width=8cm]{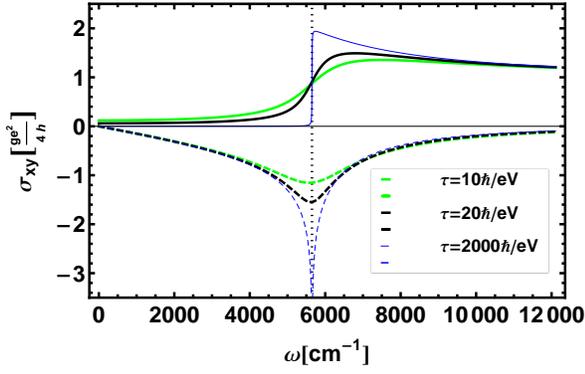}
\caption{\label{sxx_tt}
(Color online) Real (solid) and imaginary (dashed) optical conductivity versus frequency for different relaxation times which is independent of density (universal) for $\Sigma_n=0.35eV>\mu=0.1eV$. The vertical dotted lines indicate the interband threshold $\hbar \omega =2\Sigma_n$.}
\end{figure}

If the effective Zeeman field becomes larger than the chemical potential (Fermi energy) than the conductivity is exclusively due to intraband transitions and independent of density according to (\ref{sigB0})
\ba
\sigma^B_{xx}(\Sigma_n>\mu)=&{e^2\over 8 \pi \hbar}\left [
{4\Sigma_n i\over \hbar (\omega+{i\over \tau})}
\right . \nonumber\\ 
+& \left .{i\over \pi} 
\left(1+{4\Sigma_n^2\over \hbar^2(\omega+{i\over \tau})^2}\right )\ln{2\Sigma_n-(\omega+{i\over \tau})\over 2\Sigma_n+(\omega+{i\over \tau})}
\right ]
\label{sigBs0}
\end{align}
seen in figure \ref{sxx_tt}. The real part of the conductivity starts at the threshold $\hbar \omega=2 \Sigma_n$ accompanied by a minimum in the imaginary part. The finite scattering smears this step-like behavior and leads in the strong scattering limit to a constant real conductivity of universal value. The astonishing fact is that not only an universal value appears for small densities due to the chiral nature of particles but that a whole universal optical conductivity appears independent of density and solely determined by strong effective Zeeman fields.

\begin{figure}
\includegraphics[width=8cm]{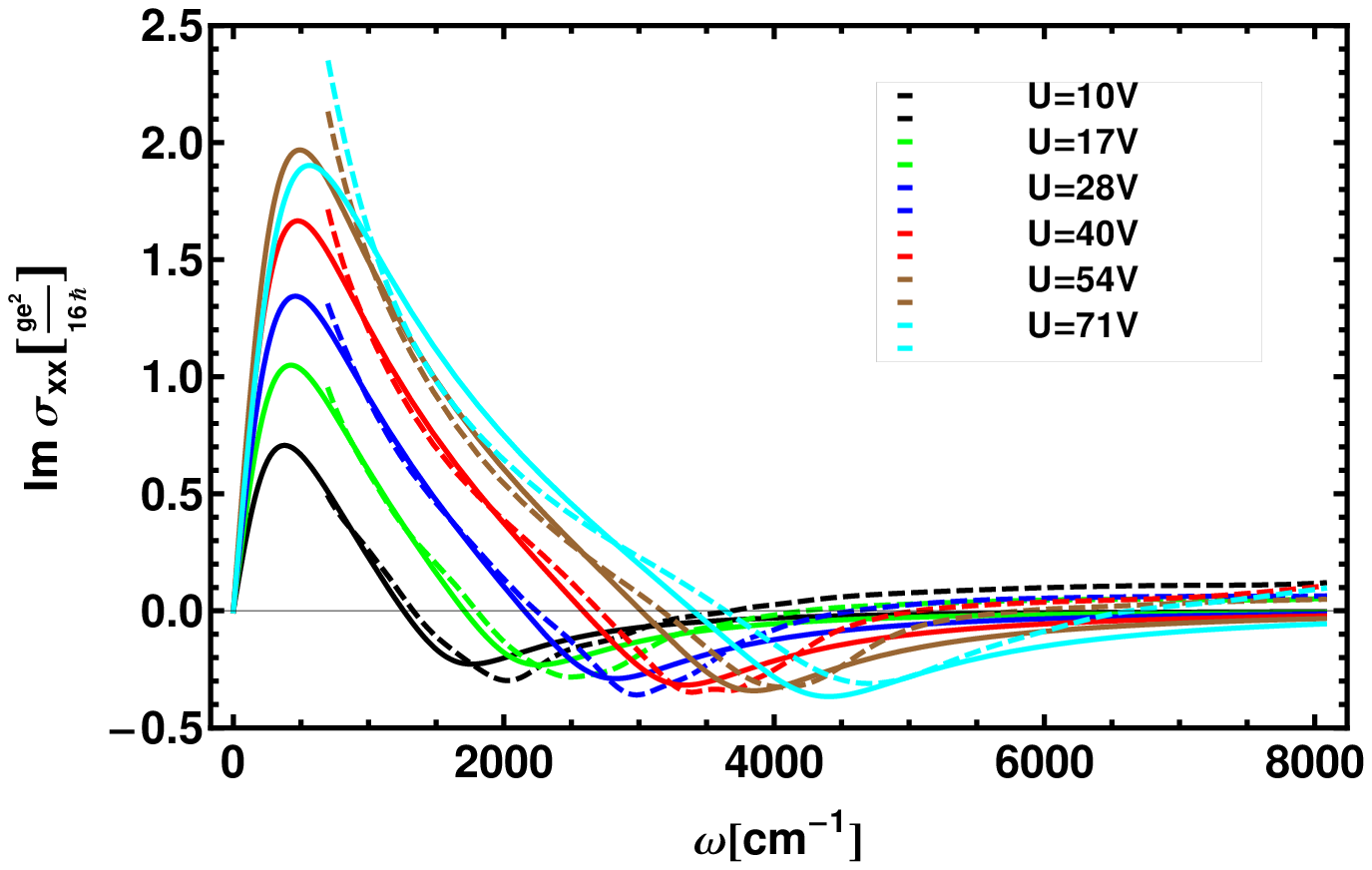}
\includegraphics[width=8cm]{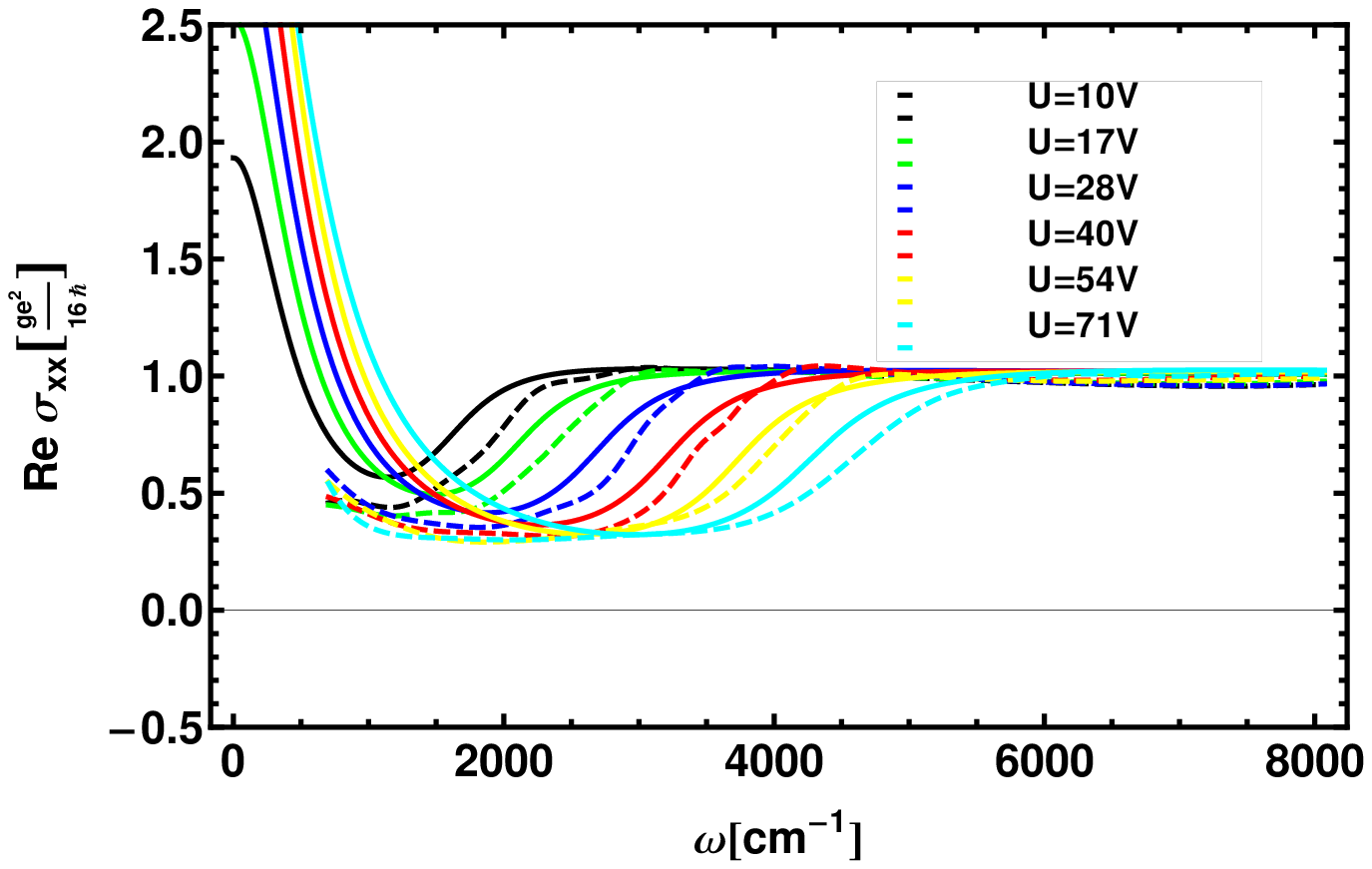}
\caption{\label{sex}
(Color online) Real (above) and imaginary (below) optical conductivity versus frequency for different applied voltages compared with the experimental data (dashed) of \cite{LHJHMKS08}. The parameters are given in figure \ref{sxx_w}.}
\end{figure}

In figure \ref{sex} we compare with the experimental data of \cite{LHJHMKS08}. We find the best fit with the help of fitting the relaxation time and the effective Zeeman field presented in the next figure~\ref{st_fit}. One sees that the relaxation time decreases with increasing density and the effective Zeeman field increases. The found relaxation time is in agreement with the theoretical value of screened charged impurities of density $n_s$. The relaxation time reads \cite{NMA06}
\be
{\hbar \over \tau}={n_s k_f\over 2 \pi \hbar^2 v} \int\limits_0^{2 \pi} d\theta V^2(q) (1-\cos{\theta}){1+\cos{\theta}\over 2}
\label{t}
\ee
with $\mu=v k_f$ and the scattering wave vector $q=2 k_f \cos{\theta/2}$. The angular expressions come from the wave-vector dependence of the relative phase of graphene Bloch band wave functions \cite{NMA06}. Since $q<2 k_f$ we have for the polarization \cite{Mc02}
$
\Pi (0,q)=-{\partial n\over\partial \mu}
$
and for the dielectric function
\be
\epsilon(0,q)=1-V(q)\Pi(0,q)=1+\alpha_V {g k_f\over 2 q}
\ee
with the fine structure constant of graphene $\alpha_V=e^2/4\pi\epsilon_0 \hbar v$ and the degeneracy $g=4$. The screened Coulomb potential in quasi 2D reads therefore\cite{Mc02}
\be
V(q)={e^2\over 2 \epsilon_0} {\hbar\over q+\alpha_V {g k_f\over 2 }}.
\ee 
Using this potential in (\ref{t}) one gets
\be
{\tau \mu \over \hbar}={8\over \pi g \alpha_V^2} {n\over n_s}\, c\left ({\alpha_V g\over 4}\right )
\label{relax}
\ee
with the constant
\ba
&c(a)={8\over 1 \!+\! 2 a^2 (9\! -\! 60 a^2 \!+\! 56 a^4) \!+\! 
    16 a^3  (4 \!-\! 7 a^2)\sqrt{a^2 \!-\! 1}}.
\end{align}
For the unscreened potential we have $c(0)=8$ and for the screened one $c(\alpha_V g/4)\approx 74$ such that the the relaxation time would be 9 times smaller for unscreened potentials. The relaxation time for the screened potential is also plotted in figure \ref{st_fit}. Here we have assumed that the ratio of density to impurities $n/n_s\approx 1/4$ according to valley degeneracy and charge neutrality. It is visible that the unscreened potential is ruled out and the relaxation times agrees with the best-fitted one. If we had used the analytical result (\ref{relax}) directly and fitted only the effective Zeeman field, visibly the same figure \ref{sex} would have appeared. This shows that the actual relaxation time from impurity concentration is quite realistic.

\begin{figure}
\includegraphics[width=8cm]{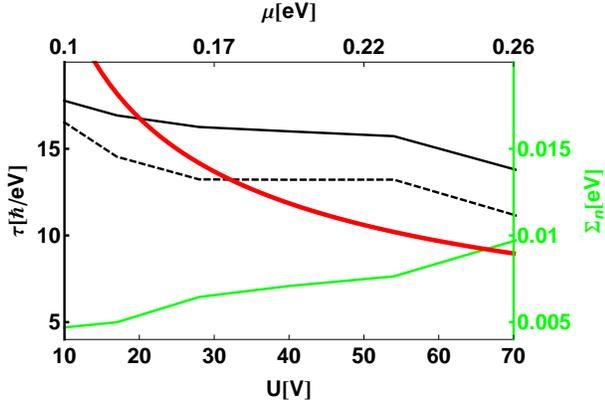}
\caption{\label{st_fit}
(Color online) Fitting effective Zeeman field and relaxation time to reproduce the experimental data in figure \ref{sex}. The best fit only with relaxation time (dashed) is compared to the relaxation time for screened Coulomb impurities (thick black) together with the analytical (red thick) result of (\ref{relax}). The effective Zeeman field (green) is given with the units on the right axis. }
\end{figure}

In the theoretical approach\cite{SPN08}  it was found that the conductivity is still underestimated even taking into account the effect of optical and acoustic phonons as well as charged impurities and midgap states. We find a slight overestimate when allowing an effective Zeeman field.

To illustrate the relevance of the effective Zeeman field we plot in the next figure \ref{st_fitU} the curves with and without Zeeman field. We find a fairly good description of the experimental optical conductivity with a simple relaxation time assuming only charged impurities. The Zeeman field leads to a better agreement of the dissipative part of conductivity with experiments.

\begin{figure}
\includegraphics[width=8cm]{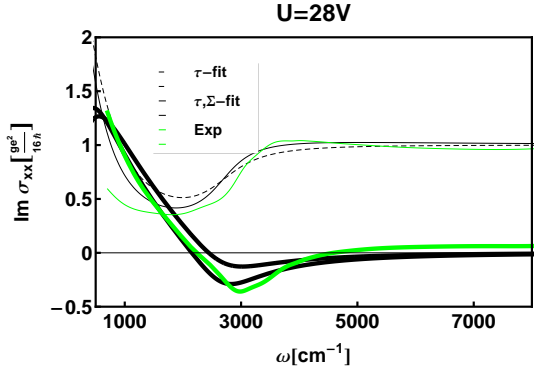}
\caption{\label{st_fitU}
(Color online) Comparison of best fit to real (thin) and imaginary (thick) experimental data (green) of figure \ref{sex} for $U=
28
V
$ 
using once the relaxation time and the effective Zeeman field (solid) and once only the relaxation time (dashed).}
\end{figure}

\begin{figure}
\includegraphics[width=8cm]{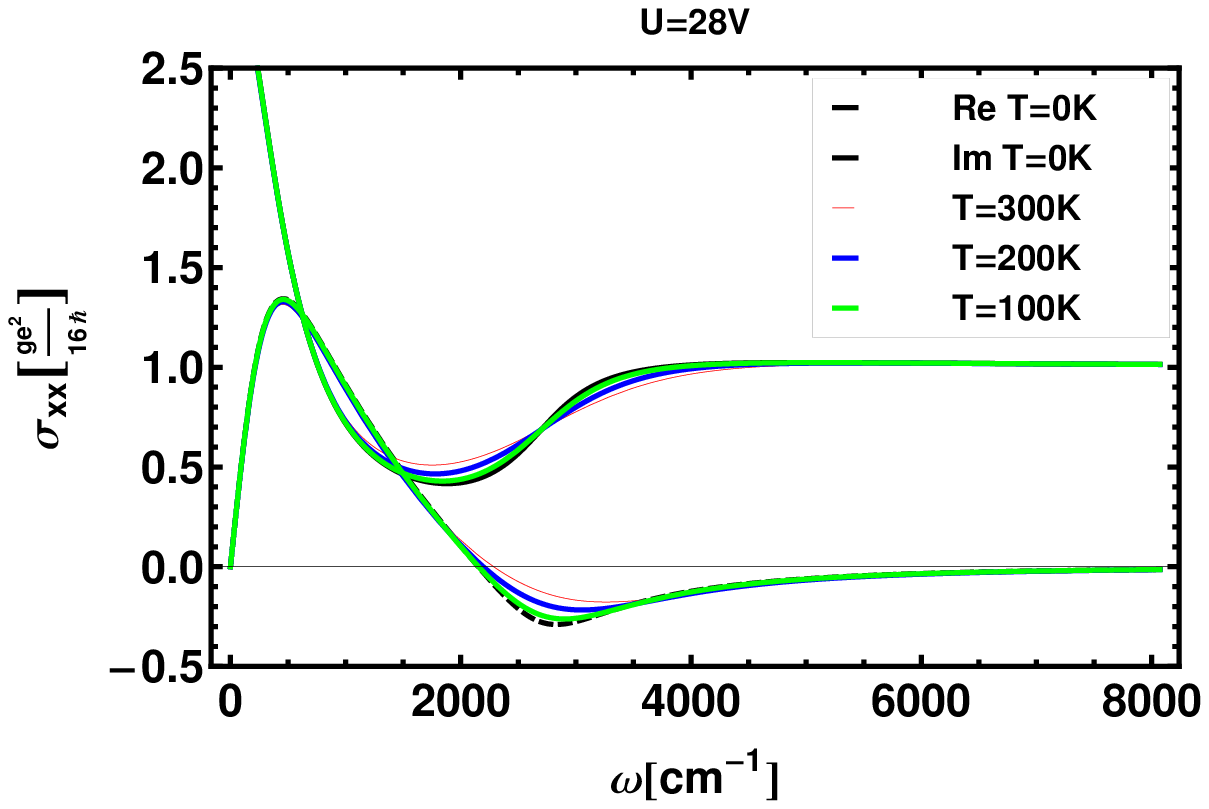}
\includegraphics[width=8cm]{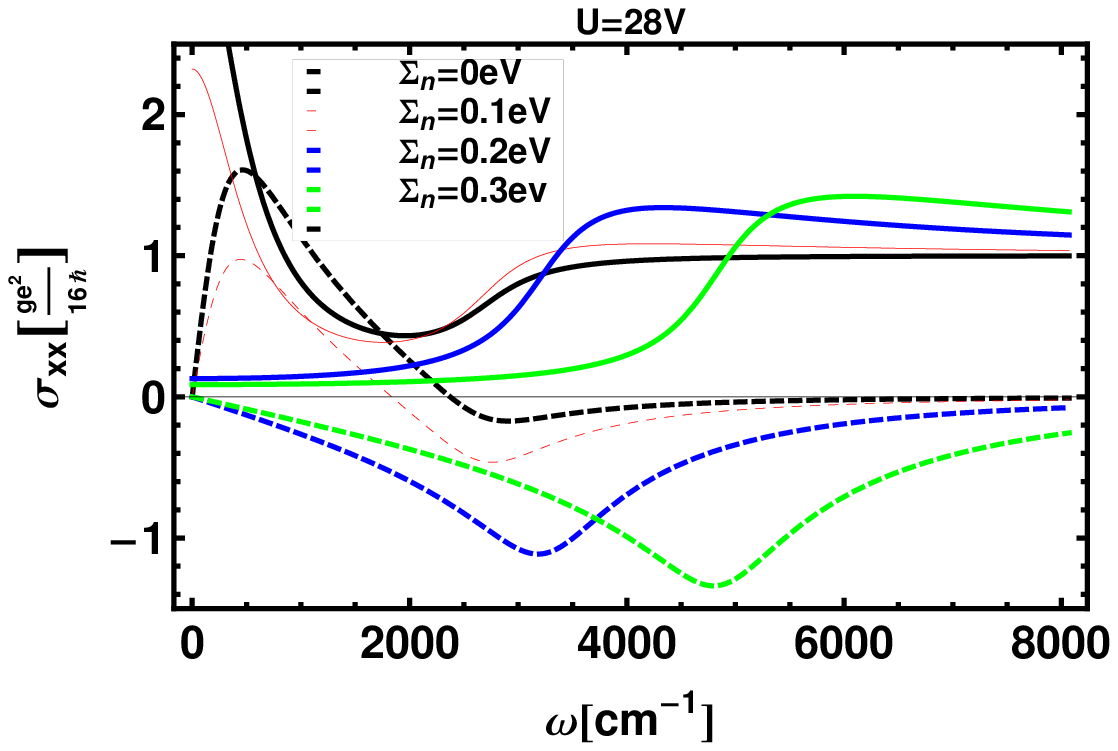}
\caption{\label{st_T}
(Color online) The influence of finite temperatures (above) and the effective Zeeman field (below).}
\end{figure}

The influence of finite-temperatures is marginal as seen in figure \ref{st_T} while the effective Zeeman field leads to appreciable modifications. The extension of the optical conductivity including the gap can be found in \cite{GSC06}. Here we consider only the quasiclassical approximation which results into an effective Zeeman field and therefore only the first quantum-Hall level which can be extended \cite{GSC07}.

\subsection{Hall conductivity}
Let us assume the magnetic field in $z$-direction and the electric field in $x$-direction. Then we have for graphene
\ba
&\sigma_{xy}
=2 e^2 \lambda^2 {\Sigma_n\over v} {\cal E}
\nonumber\\
&={e^2\over 2 \pi \hbar} \!\left({\Sigma_n\over T}\right) \!\left ({T\tau_\omega\over\hbar}\right)^2 \!\!\int\limits_{\Sigma_n\over T}^\infty \!\!d x{
\left({\rm e}^{-x\!-\!\mu\over T}\!+\!1\right )^{\!-1}\!\!\!-\!\left ({\rm e}^{x\!-\!\mu\over T}\!+\!1\right )^{\!-1}
\over 1+4 x^2 \left ({T\tau_\omega\over \hbar}\right )^2}. 
\label{sxy}
\end{align}
The Hall conductivity for particles with $p^2/2m$ dispersion at zero temperature  and $\mu>\Sigma$ reads \cite{M15}
\ba
&\sigma_{xy}={e^2\over 4 \pi \hbar^2} \Sigma_n \tau_\omega \arctan{\left [
{4 \epsilon_v \tau_\omega/\hbar\over \hbar^2+4 (2 \epsilon_v \mu+\Sigma_n^2)\tau_\omega^2}
\right ]}
\label{146}
\end{align}
with the energy $\epsilon_v=mv^2$ and the chemical potential (Fermi energy) $\mu$. The dynamical result is given by the frequency dependence (\ref{tw})
in (\ref{146}) and (\ref{sxy}). We see that the anomalous Hall effect vanishes with vanishing effective Zeeman field (\ref{147}). The latter one being an external magnetic field or an effective magnetized domain.

The infinite-mass limit reads now
\be
\sigma_{xy}&=&{e^2\over 4 \pi \hbar^2} \tau_\omega\Sigma_n\, \left \{ \begin{matrix}
 {\rm arccot} {2\mu \tau_\omega\over \hbar},
&\mu>\Sigma_n
\cr
{\rm arccot} {\Sigma_n \tau_\omega\over \hbar},
&\mu<\Sigma_n
\end{matrix}
\right .\nonumber\\\nonumber\\
&\to&
{e^2\over 8 \pi \hbar}\left \{
\begin{matrix}
{ \Sigma_n\over \mu}+o(\tau_\omega^{-1}) ,\quad \mu>\Sigma_n
\cr
1+o(\tau_\omega^{-1}),\quad \mu <\Sigma_n
\cr
{\Sigma_n \tau\over \hbar} \left ( \pi-{4 \tau \mu\over \hbar}+o(\mu^2)\right ),\quad \mu>\Sigma_n
\end{matrix}\right .
\label{sigHall0}
\ee
which is nothing but the zero-temperature limit of (\ref{sxy}). One sees that a step-like structure appears if the chemical potential exceeds the effective magnetic field $\Sigma_n$. Remarkably, for larger effective Zeeman field $\Sigma_n$ we obtain a result independent of the chemical potential and therefore independent on the density. Even in the limit of vanishing scattering the universal value 
$
\sigma_{xy}\to {e^2/ 8\pi \hbar}
$
appears.

\begin{figure}
\includegraphics[width=8cm]{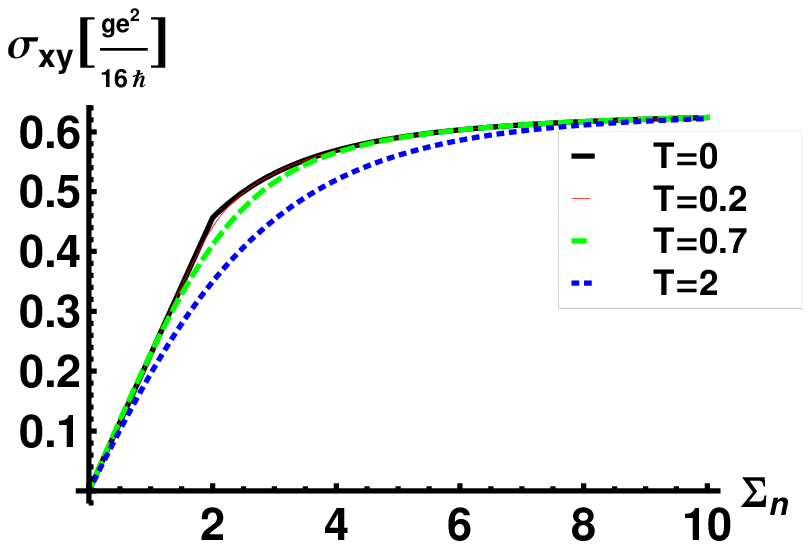}
\includegraphics[width=8cm]{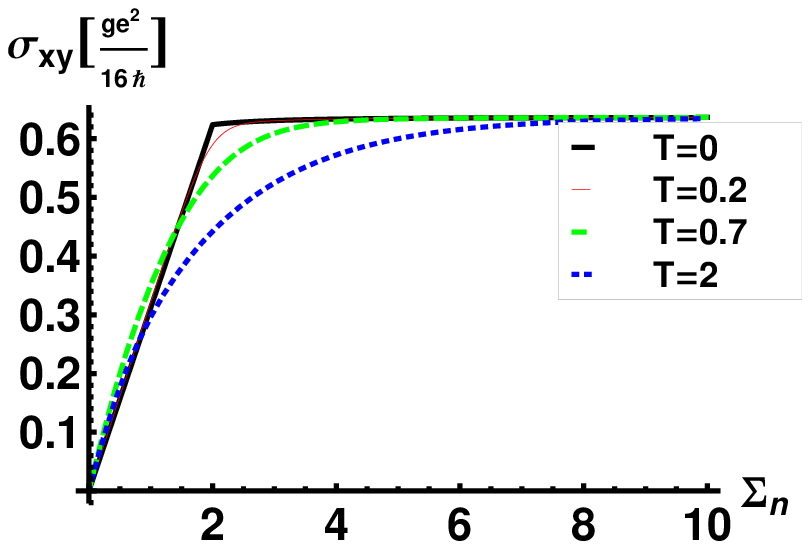}
\caption{\label{sxy_m2tt1_sT}
(Color online) Temperature dependence (in arbitrary energy units EU) of the static Hall conductivity (\ref{sxy}) versus effective Zeeman field with $\mu=2 EU$ for $\tau=0.2\hbar/EU$ (above) and
$\tau=1 \hbar/EU$ (below) where the universal limit in these units is $2/\pi$.}
\end{figure}

\begin{figure}
\includegraphics[width=8cm]{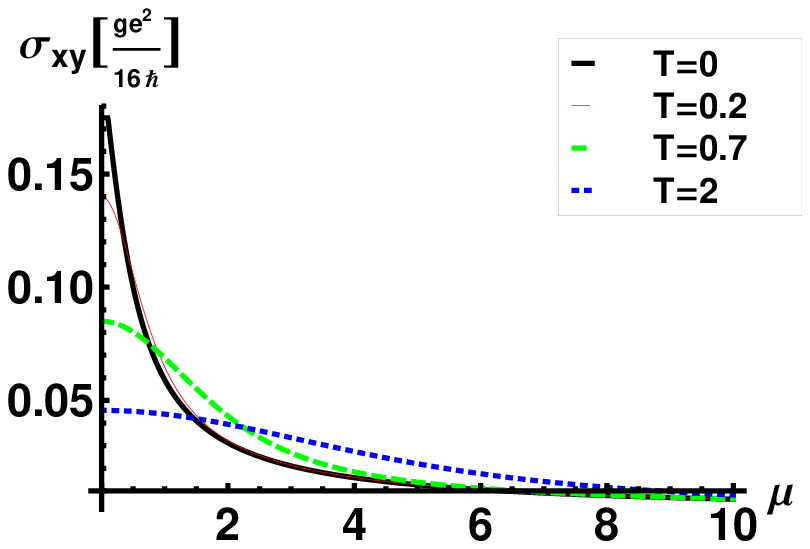}
\includegraphics[width=8cm]{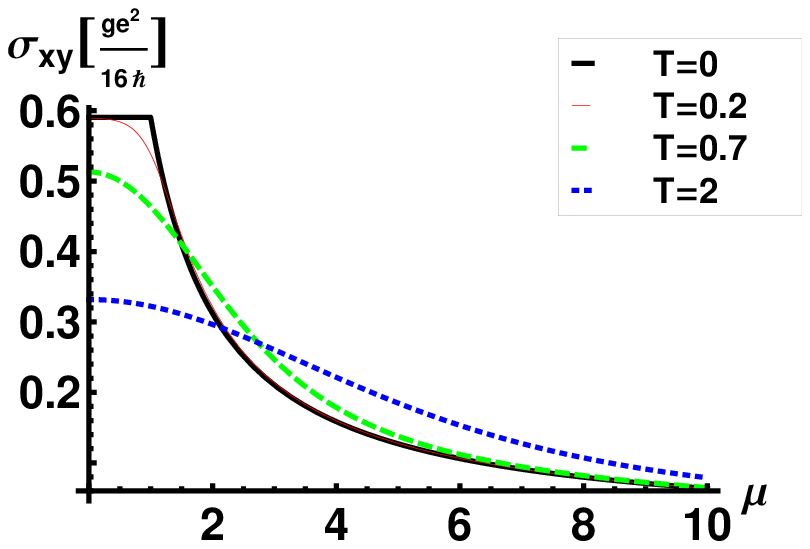}
\caption{\label{sxy_s01_T}
(Color online) Temperature dependence (in arbitrary energy units EU) of static Hall conductivity (\ref{sxy}) of graphene with $\Sigma_n=0.1 EU$ (above) and $\Sigma_n=1 EU$ (below) and $\tau=1 \hbar/EU$.}
\end{figure}

 The temperature dependence of the static Hall conductivity is seen in figure \ref{sxy_m2tt1_sT}. For large effective Zeeman fields the universal limit is approached. The finite-temperature results versus chemical potential are seen in figures \ref{sxy_s01_T} which shows the appearance of the threshold at chemical potentials in the order of the Zeeman energy above which the conductivity decreases while below it is constant and determined uniquely by the Zeeman energy.

\begin{figure}
\includegraphics[width=8cm]{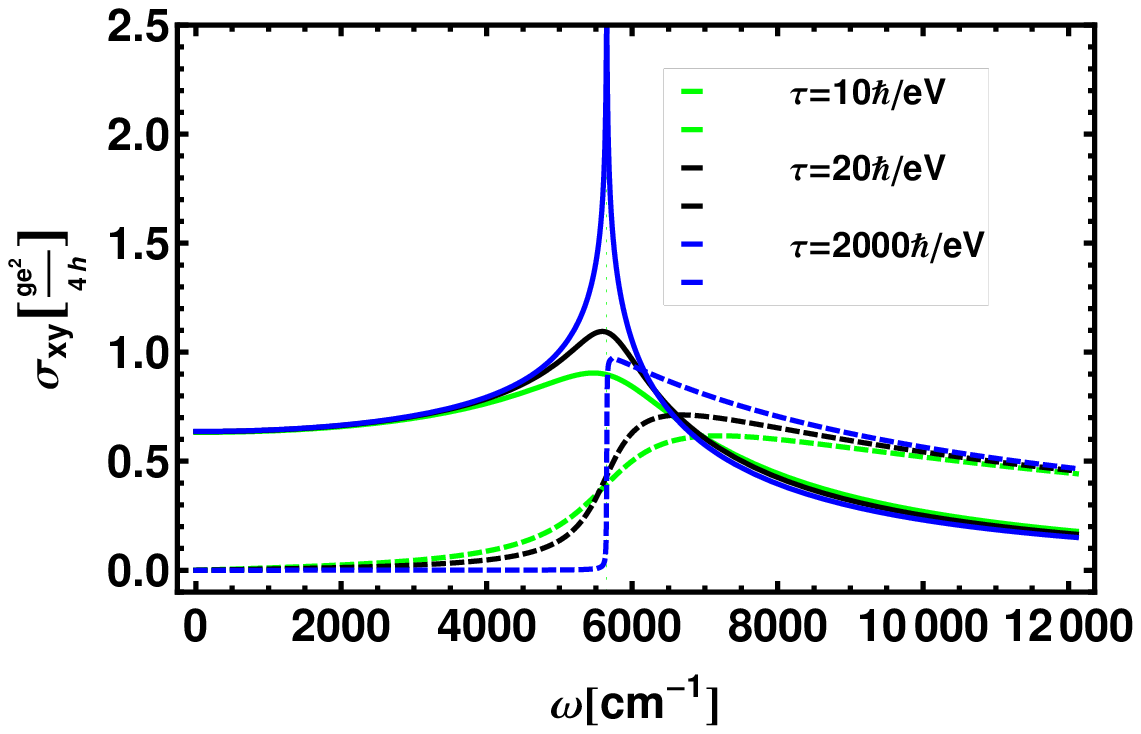}
\includegraphics[width=8cm]{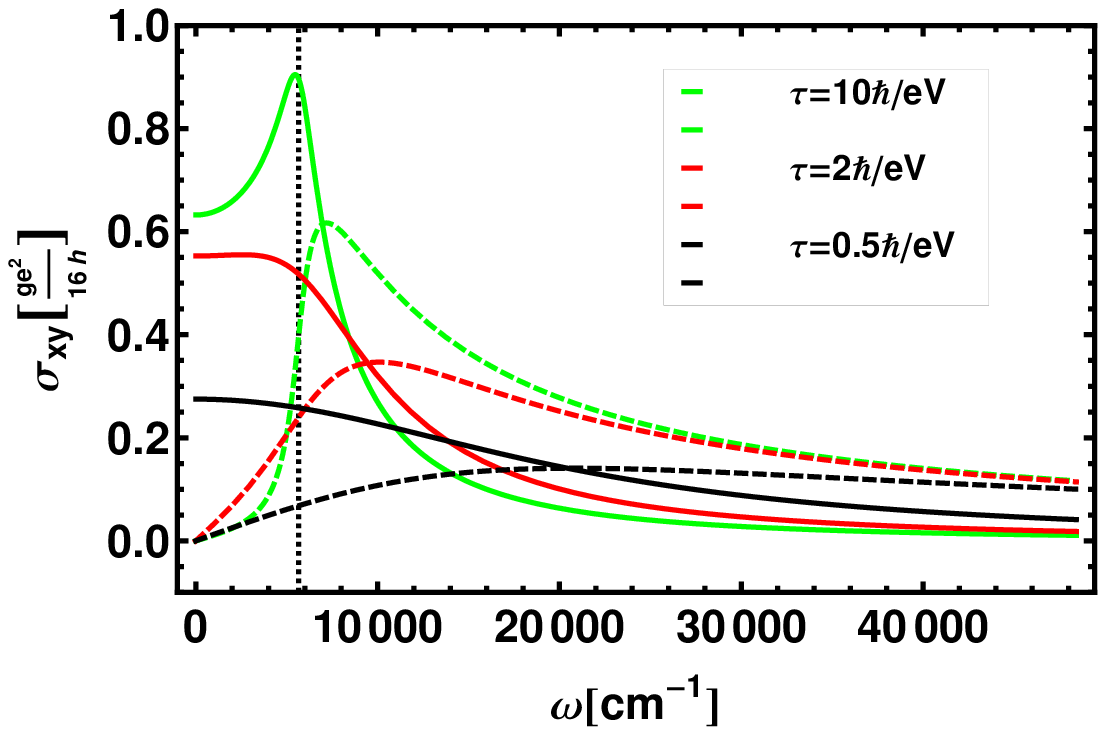}
\caption{\label{sxy_tt_w}
(Color online) Dynamical Hall conductivity (\ref{sigHall0}) of graphene with $\Sigma_n=0.35 eV$ and $U=10V$ for different relaxation times. The dashed vertical line gives the interband transition energy $max[2\mu, 2\Sigma_n]$}
\end{figure}

The dynamical conductivity according to (\ref{tw}) is shown in figure \ref{sxy_tt_w}. The real part has a sharp maximum at the interband transition frequency where the imaginary part starts. This feature is smeared out by collisions.

\subsection{Expansion in fine structure constant}

The different conductivities can be expanded in terms of the graphene fine structure constant $\alpha_V$. This expansion is also not interchangeable with the small-density expansion. The results are identical  with and without screening in lowest order in $\alpha_V$ and read
\be
\sigma_{xy} ={g e^2\over 16 \hbar } \left \{ \begin{matrix}
{\Sigma_n\over \mu}\left ({2\over \pi}-{\pi \over 96}{n_s\over 4 n}\alpha_V^4+o(\alpha^6) \right ),\quad \mu>\Sigma_n
\cr
{2\over \pi}-{\pi \mu^2 \over 96\Sigma_n^2}{n_s\over 4 n}\alpha_V^4+o(\alpha^6),\quad \mu<\Sigma_n
\end{matrix} \right ..
\ee
The second line appears as well if we first expand up to orders $o(\mu^2)$
which agrees with the universal limit.

For the longitudinal conductivity we obtain
\be
\sigma_{xx} ={g e^2\over 16 \hbar } 
\left \{ \begin{matrix}
{64 (\mu^2-\Sigma_n^2)\over \mu^2\pi^2}{4 n\over n_s}\alpha_V^{-2}+o(\alpha_V^0) ,\quad \mu>\Sigma_n
\cr
{\mu\over 12 \Sigma_n\pi^2}{n_s\over 4 n}\alpha_V^{2}+o(\alpha_V^4),\quad \mu<\Sigma_n
\end{matrix} \right .
\ee
which is in contrast to the expansion first in density and then in fine structure constant
\be
\sigma_{xx} ={g e^2\over 16 \hbar } \left (1+{8162 \mu^6 \over 3 n_s^2\pi^7 v^6}\alpha_V^{-6}+o(\mu^7)+o(\alpha_V^{-4}) \right )
\ee
which agrees with the universal limit.
One sees completely different expansion schemes depending whether we expand first with respect to the density or with respect to the fine structure constant.

\section{Pseudospin conductivity\label{spincurrent}}

Now we consider the pseudospin current according to (\ref{current}) 
\ba
\V S_j&=2 \sum\limits_p\left (\partial_{p_j}\epsilon\delta \V g\!+\!\delta f \partial_{p_j} \V b \right )
\label{current1}
\end{align}
where the first part is the normal and the second part is the anomaly part.
For graphene in the limit of infinite mass we would not expect a normal part since $\partial_{p_j}\epsilon=p_j/m\to 0$ at a first glance. However there is a subtle problem here. In \cite{M15} we have calculated this first normal part as 
\be
\V S_\alpha=-{e \tau \over m_e(1-i\omega \tau)}\V s E_\alpha+\V \sigma_{\alpha\beta} E_\beta.
\ee
The normal spin-Hall coefficient consists analogously as the anomalous Hall effect of a symmetric and an asymmetric part ($\omega\to \omega+i/\tau$)
\be
\left .
\begin{matrix}
\V \sigma_{\alpha\beta}^{\rm as}
\cr\cr
\V \sigma_{\alpha\beta }^{\rm sym}
\end{matrix} 
\right \}={e\over m_e\omega} \sum\limits_p {p_\alpha g  \over
  1-{\omega^ 2\over 4|\Sigma|^2}}\,\left \{ \begin{matrix}
{i\omega\over
    2|\Sigma|} \V e\times \partial_{p_\beta} \V e
\cr\cr
i\partial_{p_\beta} \V e
\end{matrix} \right .
\label{spinHall}
\ee
with the explicit integration in zero temperature and linear Rashba coupling\cite{M15}
\ba
&\sigma_{yx}^z\!=\!{e\over 8 \pi \hbar}\!\left [
1\!-\!{\hbar^2\!+\!4 \Sigma_n^2 \tau_\omega^2\over 4 \epsilon_v \tau_\omega \hbar} 
\arctan{\!\left (\!
4\hbar \epsilon_v \tau_\omega\over \hbar^2\!+\!4 \tau_\omega^2 (2 \epsilon_v \mu\!+\!\Sigma_n^2)
\!\right )}
\!\right ]
\nonumber\\
&\sigma_{xx}^z={2\over \hbar} \Sigma_n \tau\sigma_{yx}^z
\label{SHR}
\end{align}
with $\tau_\omega= \tau/(1-i \omega \tau)$ and $\epsilon_v=m v^2$. Neglecting the selfenergy and using the static limit it is just the result of \cite{Schliemann:2004,Sch06}. 
For the Dresselhaus linear spin-orbit coupling one has (\ref{SHR}) with opposite sign.

For graphene we perform the limit of infinite mass of (\ref{SHR}). Amazingly we obtain just the universal limit 
\be
\lim\limits_{m\to \infty}\sigma_{yx}^z={e\over 8\pi \hbar}
\ee
contrary to the expectation at the beginning of this chapter that this normal parts of pseudospin current should vanish. This shows the highly nontrivial transition from quadratic dispersion with spin-orbit coupling towards the Luttinger-type dispersion of Dirac particles. The infinite-mass limit of the normal spin-Hall conductivity leads to a finite result though the quasiparticle velocity $\partial_{p_j}\epsilon$ vanishes and we would have expected no result from the normal current. One obtains such a zero effect if one performs first the limit of vanishing density and then the absence of collisions $\tau \to \infty$. The same is true if we first perform the limit of vanishing collisions and then the static limit. We illustrate some orders of limits in table~\ref{tab3}. The conclusion is that the limits of infinite mass to let the quasiparticle velocity vanish cannot be interchanged with the static limit.

\begin{table}
\begin{tabular}{|c|c|c||c|}
\hline
\multicolumn{3}{|c||}{order of limits}& $\sigma_{yx}^z=\zeta {e^2\over 8 \pi \hbar}$\\
\hline
3.&2.&1.& $\zeta$
\cr
\hline 
\hline 
&&$m\to\infty$& $ 1$
\cr
\hline
&$m\to\infty$ &$\omega\to 0$& $1$
\cr
\hline
$m\to\infty$&$\Sigma\to 0$ &$\omega\to 0$& $ 1$
\cr
\hline
$m\to\infty$&$\mu\to 0$ &$\omega\to 0$& $ 1$
\cr
\hline
$\tau\to\infty$&$\mu\to 0$ &$\omega\to 0$& $ 0$
\cr
\hline
&$\omega\to 0$&$\tau\to\infty$& $0$
\cr
\hline
&$m\to\infty$ &$\tau\to\infty$& $1$
\cr
\hline
\end{tabular}
\caption{\label{tab3} Different orders of limits in $\sigma_{yx}^z$.}
\end{table}


The universal constant $e/8\pi\hbar$ has been first described by
\cite{SCNSJD04} and raised an intensive discussion. It was shown that the
vertex corrections cancel this constant \cite{IBGML04,RS05}. A suppression of
Rashba spin-orbit coupling has been obtained due to disorder \cite{CL05}, or
electron-electron interaction \cite{D05} and found to disappear in the
self-consistent Born approximation \cite{Liu06}. The conclusion was that the
two-dimensional Rashba spin-orbit coupling does not lead to a spin-Hall effect
as soon as there are relaxation mechanisms present which damp the pseudospins
towards a constant value. Beyond the meanfield, in relaxation-time approximation by including vertex
corrections \cite{ZNA01} or treating collision integrals \cite{SMEH06, SSS12} lead to vanishing spin Hall current. The spin-Hall effect does not vanish with magnetic fields or spin-dependent scattering processes \cite{Sch06}. This was discussed by kinetic theory \cite{GSDR08} for dirty and clean regime and the Kubo formula \cite{IKIIBW06} where it was shown that the Zeeman field gives a nonvanishing spin Hall current. In \cite{BRV11} magnetic impurities were treated in tight-binding approximation and the importance of the order of large scattering time and small frequencies were pointed out.

Please note that the universal constant in (\ref{SHR}) is necessary to obtain the correct small spin-orbit coupling result
\be
\sigma_{yx}^z={e\over \pi \hbar} {\mu \tau^2\over (1-i \omega \tau)^2+4 \Sigma_n^2 \tau^2} \epsilon_v+o(\epsilon_v^2).
\ee
Without the Zeeman term $\Sigma_n\to 0$ and for small spin-orbit coupling this
agrees with the dynamical result of \cite{BK06} where the definition of pseudospin
current has been employed in terms of physical argumentation. Again the result
here differs from the resonant structure found in \cite{BK06} by the $arctan$ term but the static limit agrees with the result of \cite{Schliemann:2004,Sch06}. 
The dynamical density-density response of graphene beyond the Dirac cone approximation can be found in \cite{SSP10} and the dielectric properties including spin-orbit interactions are discussed in \cite{SSS12}. An overview about different scattering mechanisms can be found in \cite{Pe10}.

We consider now the anomalous part as the second part of (\ref{current1}). Calculating explicitly for graphene with $\V b=v (p_x,p_y,0)$ one obtains
\be 
\V S_j^a=2 \sum\limits_p \delta f \partial_{p_j} \V b=2 v \delta n \V e_j
\ee
which shows that the pseudospin current follows the spinor direction. From the mean-field-free response functions (\ref{res}), as it should be for conductivity, we obtain the density variation
\be
\delta n={v\over \omega} \V q\cdot \delta \V s
\ee
and using (\ref{deltarho}) one gets after performing the angular integration
\be
\V q\cdot \delta \V s=e \V q \cdot \left [-{i\over \bar \omega} ({\cal D}+{\cal B}) \V e_\alpha+2 {\Sigma_n\over \bar \omega^2}{\cal E} \V e_z\times \V e_\alpha\right ] E_\alpha.
\ee
Using the electric field again in x-direction the anomalous pseudospin current conductivity reads
\be
\V \sigma_{j x}^a={v\over \omega} \V e_j \left [{\sigma_{xx}\over e} q_x+{\sigma_{xy}\over e} q_y\right ]
\ee
with the anomalous current conductivities (\ref{con1}) and (\ref{sxy}) and the two possible directions $j=x,y$ providing pseudospin-Hall and pseudospin conductivity. We see  that the anomaly pseudospin conductivity is wavelength dependent and vanishes in the long-wavelength limit.

\section{Dielectric function}

The induced density is related to the external potential by the response function $\delta n=\chi \Phi$. The screened potential $V^s$ is given by the external potential $\Phi$ and the interacting one $V_0$ by $V^s=V_0\delta n+ \Phi$. The ratio of the screened to the external potential is the inverse dielectric function
\be
{1\over \epsilon}={V^s\over \Phi^{\rm ext}}=1+V_0\chi.
\ee
In other words, the electrons feel the effective potential
$V_0/\epsilon$. Though the Coulomb interaction is repulsive a change in sign
of the dielectric function indicates an effective attraction as it is the mechanism for Cooper pairing due to phonon coupling. It is interesting to search for such regions as possible range where Cooper pairing can occur in graphene.

\begin{figure}
\includegraphics[width=8cm]{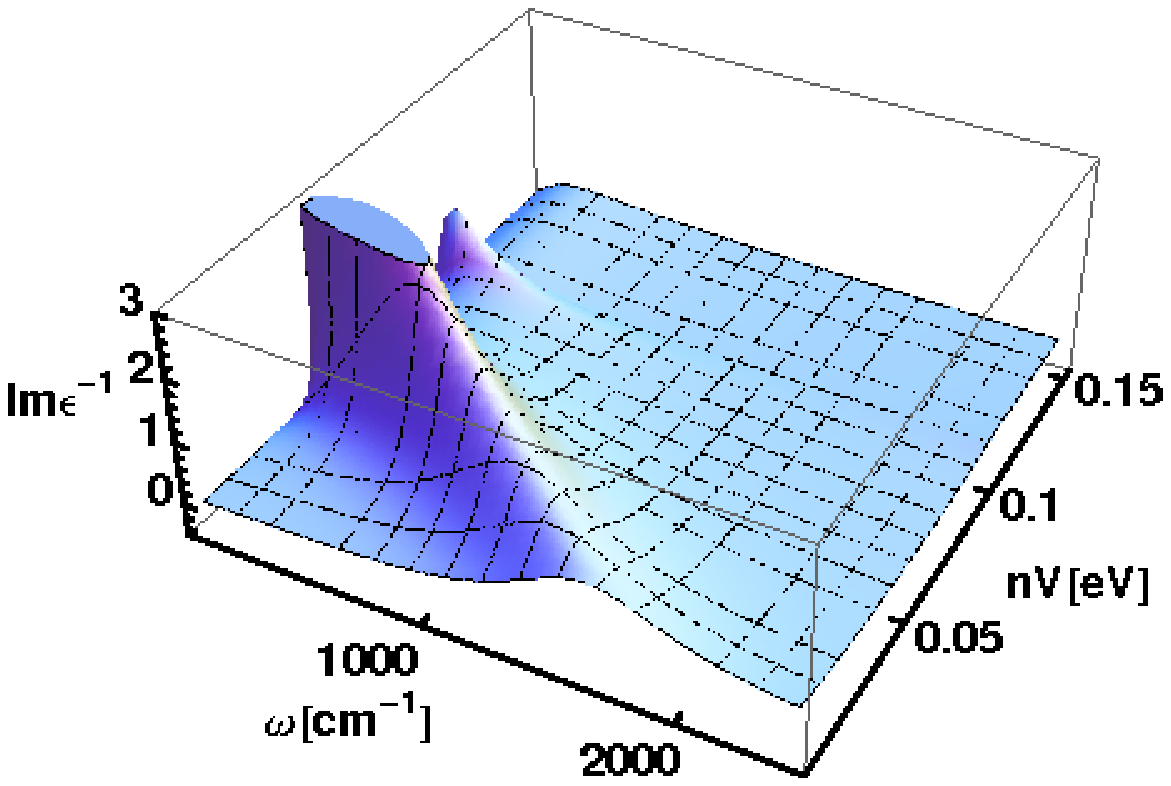}

\includegraphics[width=8cm]{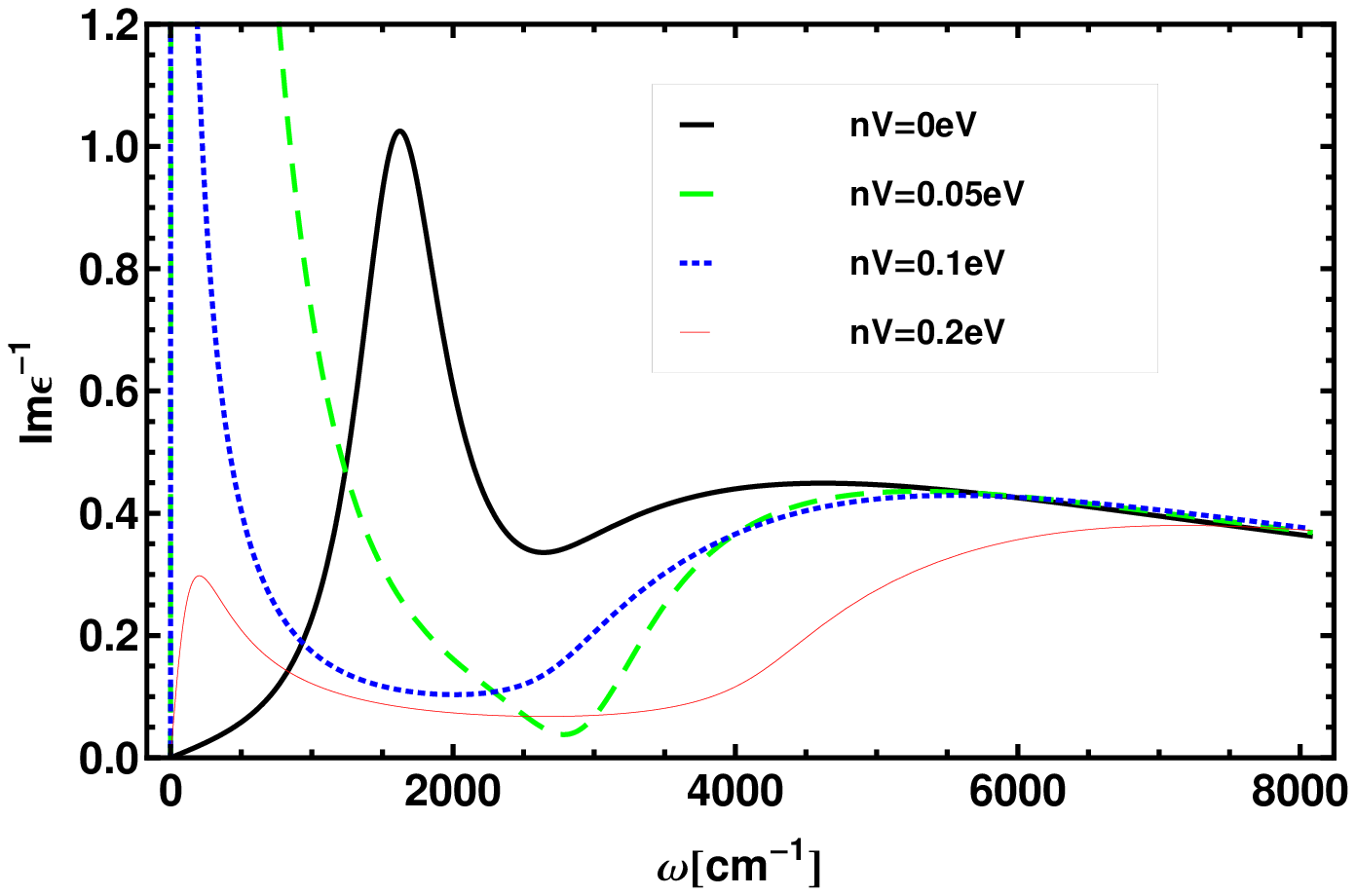}
\caption{\label{ex_wq1_28V}
(Color online) The excitation function versus frequency and domain interactions as 3D plot (above) and cut at different domains strength (below) for $v q=1eV$ and a density corresponding to $28V$ of figure \ref{st_fitU}.
}
\end{figure}

The response can be calculated straightforwardly form the linear equation
system (\ref{res}). The result is a lengthy expression. Only special
expansions are needed here. In the optical regime the coupling velocity $v$ of
graphene is about $1/300$ of the speed of light and one has the small parameter
\be
\eta={v q\over \hbar \omega}={1 \over 300}. 
\ee
Expanding the response function one obtains in the optical regime
\be
{1 \over \epsilon}=1-{\omega\over \bar \omega}\eta^2-i {\omega^2\over {\bar \omega}^2 \bar \sigma} {\pi \alpha\over 16} \eta^3+o(\eta^4)
\ee
with the dimensionless conductivity $\bar \sigma=16 \hbar \sigma_{xx}/ge^2$ from (\ref{con1}) and the fine structure constant $\alpha\approx1/137$. We see that the deviation of the inverse dielectric function in the optical regime is only in orders of $10^{-4}$ and cannot lead to any sign change.

This is different if we consider the non-optical regime, where $\hbar w\ne c
q$. Then $\eta$ is no small parameter anymore and we can only expand with
respect to the fine structure constant leading to
\be
\epsilon=1+i{\bar \sigma \over 1+ {2 \Sigma_n V v^2 q^2\over \omega \hbar^4{\bar \omega}^3\omega}{\cal F}}{\pi\alpha\over 16}+o(\alpha^2)
\ee  
where ${\cal F}$ from (\ref{const}) reads for $T=0$
\be
{\cal F}=-{\Theta (\mu-\Sigma_n)\over 4 \pi  \hbar^2v^2} {\mu^2-\Sigma_n^2\over 1+4 \tau_\omega^2\mu^2}.
\ee

The collective modes are visible at places where the real part becomes zero
and a small imaginary part of $\epsilon$ representing the damping. This is best visualized by the excitation function of density fluctuations, $-{\rm Im} \epsilon^{-1}$, as plotted in figures \ref{ex_wq1_28V}. We see that with increasing magnetic domain strength $V$ the collective peak becomes sharpened and shifted towards lower frequencies. The second peak around $V=0.1$ is at the line $\mu\le \Sigma_n$ where ${\cal F}$ vanishes. In \cite{M15a} negative excitations indicating spin-separation instabilities have been found for certain polarizations. Here in graphene no such regions are observed.

\begin{figure}
\includegraphics[width=8cm]{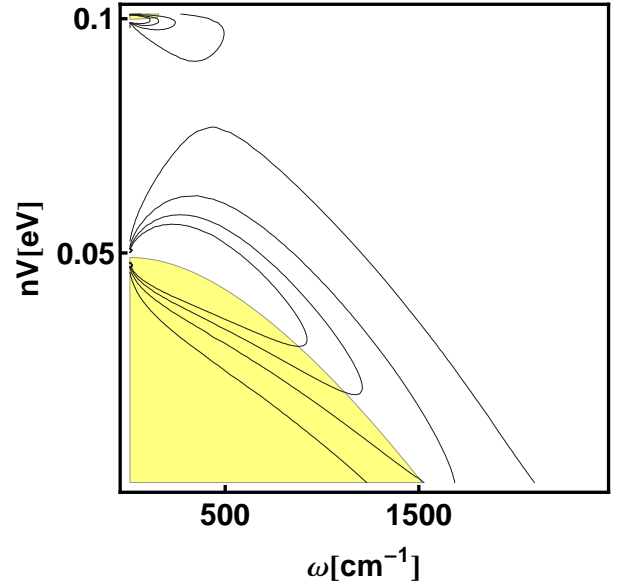}
\caption{\label{exre_wq1_28V}
(Color online) The contour lines of figure \ref{ex_wq1_28V} together with the range where the
real part of dielectric function becomes negative (yellow).}
\end{figure}

In the next figure \ref{exre_wq1_28V} we plot the contours of the excitation
functions indicating the small damping together with the area where the real
part of the dielectric function changes sign. We see that the lower part
of the collective excitation is accompanied by a negative area. Since this
comes together with small damping, i.e. with a small imaginary part of $\epsilon$, we can consider
this range as the one where the repulsive interaction changes effectively 
into attractive interaction and pairing is possible.

This range of attractive interaction is dependent on the wave vector and the
magnetic domain meanfield as illustrated in the next figure \ref{reqw}. The magnetic domain allows this sign change at smaller wavelengths though the frequency range shrinks with increasing magnetic domain strength. The here reported sign change is a prerequisite for Cooper pairing. A detailed analysis of the possibility to have superfluidity and superconductivity in graphene can be found in \cite{LOS12,GU12,GF16}.

\begin{figure}
\includegraphics[width=8cm]{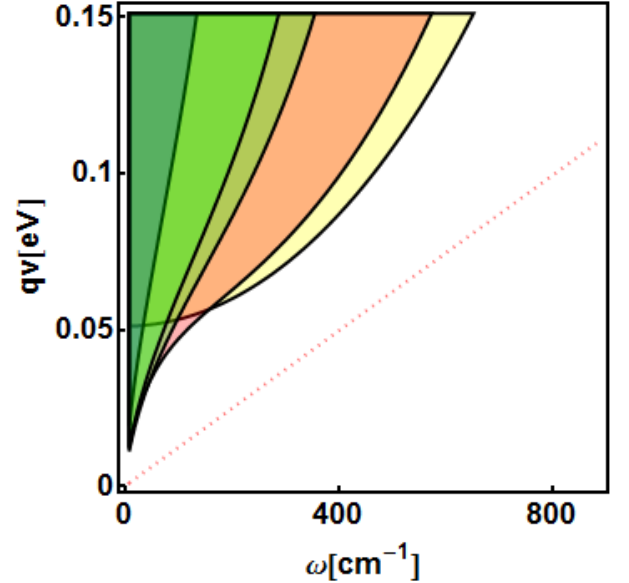}
\caption{\label{reqw}
(Color online) The range of sign change of interaction potential for different magnetic
domain meanfields; from left to right: $n V=0,0.02,0.06,0.07,0.09$eV. The
straight line illustrates $\hbar \omega= q v$.}
\end{figure}

\section{Summary}

The quantum kinetic equations for systems with SU(2) structure have been employed to describe the coupling between the Wigner distributions for charge and for pseudospin polarization. The used many-body approximation is on the level of random phase approximation (RPA). This comes about by linearizing the meanfield kinetic equation and using the relaxation time for screened Coulomb interaction.

Using the results from electrons with quadratic mass dispersion and spin-orbit coupling, one can obtain the linear Dirac dispersion of electrons in graphene by the infinite-mass limit. This allows to translate many results obtained for spin-polarized systems with spin-orbit coupling directly to graphene. We discuss here the density and pseudospin currents. For the density currents no quasiparticle part appears only anomalous ones. The anomalous currents lead to intra- and interband longitudinal conductivities and the Hall conductivity.

We have analyzed the conductivities with respect to the influence of magnetic fields and magnetic domain puddles as well as meanfields which can be recast into an effective Zeeman field. A density-independent universal behavior appears for large Zeeman fields or for small densities. The experimental optical conductivity is well reproduced by 
an intrinsic effective Zeeman field and a proper relaxation time calculated in
RPA. Various limits are found not to be interchangeable like the limit of
vanishing scattering and the static limit. Only the systematic expansion with
respect to vanishing density leads to the universal value. The pseudospin current shows a subtlety in that the infinite-mass limit leads
non-trivially to a universal value for the quasiparticle part though the
quasiparticle velocity vanishes itself. This is certainly a consequence of the nontrivial change of quadratic dispersion towards the unbounded one of Dirac particles, but the deep reason behind requires more investigations. 

The dynamical density and pseudospin response functions are derived as coupled linear equations system. By linearization of the meanfield kinetic equations for density and pseudospin, the response in random phase approximation (RPA) with relaxation time-approximation is obtained. Though the standard RPA-Lindhard polarization function vanishes for graphene, the anomalous coupling to the spin-polarization induces different forms of polarization functions which are presented explicitly. The resulting dielectric function is discussed with respect to the expansion in orders of the fine structure constant. It is found that the effective Zeeman field is enhancing and sharpening the collective mode until some critical value. This is accompanied by a frequency and wavelength range where the screened interaction changes the sign allowing the electrons to form Cooper pairs. 
Though the used level of RPA should be extended to better approximations in
order to achieve quantitative predictions for pairing, the qualitative region
is believed to maintain. Finally we should emphasize that we have considered
all quantities up to fourth order in wavelength expansion which can be
overcome by mean-field-free approximations of the polarization function
\cite{KM16}.

\bibliography{bose,kmsr,kmsr1,kmsr2,kmsr3,kmsr4,kmsr5,kmsr6,kmsr7,delay2,spin,spin1,refer,delay3,gdr,chaos,sem3,sem1,sem2,short,cauchy,genn,paradox,deform,shuttling,blase,spinhall,spincurrent,tdgl,pattern,zitter,graphene}
\bibliographystyle{aipnum4-1}
\end{document}